\documentclass[pdftex]{article}
\usepackage[top=1truein,bottom=1truein,left=1truein,right=1truein]{geometry}

\usepackage{graphicx}
\graphicspath{{./pic}}

\usepackage{authblk}
\usepackage{abstract}
\usepackage{amsmath}
\usepackage{amssymb}
\usepackage{amsfonts}

\usepackage{hyperref}
\usepackage{url}
\usepackage{orcidlink}

\usepackage{here}
\usepackage{tikz}
\usepackage{bm}
\usepackage{setspace}
\usepackage{enumerate}
\usepackage{arydshln}
\usepackage{caption}
\usepackage{color}
\usepackage{appendix}
\usepackage{lineno}
\usepackage{ulem}

\hypersetup{%
  colorlinks=true,
  linkcolor=blue,
  citecolor=blue,
  urlcolor=cyan
}

\begin{document}

\title{\textbf{A mesh-constrained discrete point method for incompressible flows with moving boundaries}}

\author[ab]{Takeharu Matsuda\,\orcidlink{0009-0007-6263-2766}}
\author[ab]{Satoshi Ii\,\orcidlink{0000-0002-5428-5385}\footnote{Corresponding author.\\ \textit{Email addresses:} matsuda.t.b64f@m.isct.ac.jp (Takeharu Matsuda\,\orcidlink{0009-0007-6263-2766}), ii.s.148c@m.isct.ac.jp (Satoshi Ii\,\orcidlink{0000-0002-5428-5385})}}

\affil[a]{
    {\small
      \textit{
        Department of Mechanical Systems Engineering, Graduate School of Systems Design, Tokyo Metropolitan University, 1-1 Minami-Osawa, Hachioji, Tokyo 192-0397, Japan
      }
    }
}
\affil[b]{
  {\small
    \textit{
      Department of Mechanical Engineering, School of Engineering, Institute of Science Tokyo, 2-12-1 Ookayama, Meguro-ku, Tokyo 152-8550, Japan
    }
  }
}

\date{}
\maketitle

\renewenvironment{abstract}
 {\quotation\small\noindent\rule{\linewidth}{.5pt}\par\smallskip
  {\centering\bfseries\abstractname\par}\medskip}
 {\par\noindent\rule{\linewidth}{.5pt}\endquotation}

\begin{abstract}
  Particle-based methods are a practical tool in computational fluid dynamics, and novel types of methods have been proposed.
  However, widely developed Lagrangian-type formulations suffer from the nonuniform distribution of particles, which is enhanced over time and result in problems in computational efficiency and parallel computations.
  To mitigate these problems, a mesh-constrained discrete point (MCD) method was developed for stationary boundary problems (Matsuda et al., 2022).
  Although the MCD method is a meshless method that uses moving least-squares approximation, the arrangement of particles (or discrete points (DPs)) is specialized so that their positions are constrained in a background mesh to obtain a closely uniform distribution.
  This achieves a reasonable approximation for spatial derivatives with compact stencils without encountering any ill-posed condition and leads to good performance in terms of computational efficiency.
  In this study, a novel meshless method based on the MCD method for incompressible flows with moving boundaries is proposed.
  To ensure the mesh constraint of each DP in moving boundary problems, a novel updating algorithm for the DP arrangement is developed so that the position of DPs is not only rearranged but the DPs are also reassigned the role of being on the boundary or not.
  The proposed method achieved reasonable results in numerical experiments for well-known moving boundary problems.

  \vskip.5\baselineskip

\end{abstract}


\section{Introduction}
Particle methods are a practical tool in computational fluid dynamics.
The methods use freely moving discrete points (DPs), called particles, which follow fluid flows and material motion, which makes it easy to address complex flows with largely moving and deforming boundaries.
There are two pioneering methods: the smoothed particle hydrodynamics (SPH) method \cite{gingold_SmoothedParticleHydrodynamics_1977} and moving particle semi-implicit/simulation (MPS) method \cite{koshizuka_MovingParticleSemiImplicitMethod_1996}.
Recently, various practical methods have been developed for applications in, for example, bioengineering (e.g., \cite{harrison_ModellingFluidFlow_2014,sinnott_InvestigatingRelationshipsPeristaltic_2012,tsubota_BloodCellDistribution_2022}), nuclear engineering (e.g., \cite{yoon_DirectCalculationBubble_2001,xiong_LagrangianSimulationThreedimensional_2019}), natural disasters (e.g., \cite{aslami_SimulationFloatingDebris_2023,zhang_StudyMovementCharacteristics_2023,gao_Verification3DDDASPH_2024}), and planetary science (e.g., \cite{golabek_CouplingSPHThermochemical_2018,citron_LargeImpactsEarly_2022}).
To strictly enforce consistency in numerical convergence, moving least-squares (MLS) approximation \cite{lancaster_SurfacesGeneratedMoving_1981} has been implemented for spatial differentiation that guarantees an arbitrary order of accuracy.
Several methods have been developed, such as MLS particle hydrodynamics \cite{dilts_MovingleastsquaresparticleHydrodynamicsConsistency_1999,dilts_MovingLeastsquaresParticle_2000}, the MLS reproducing kernel \cite{liu_MovingLeastsquareReproducing_1997}, and least-squares MPS \cite{tamai_LeastSquaresMoving_2014} with Neumann or Robin boundary conditions \cite{matsunaga_ImprovedTreatmentWall_2020,wang_ConsistentRobinBoundary_2020}.

The uniformity of the particle distribution is an important feature for achieving stable and efficient computations in particle methods because the local arrangement of particles influences conditions in the evaluation of spatial gradients.
Particularly, in pure Lagrangian particle methods, the nonuniform distribution of particles often occurs according to their motion \cite{oger_SPHAccuracyImprovement_2016}, and exhibits sparsity and clustering because of tensile instability \cite{swegle_SmoothedParticleHydrodynamics_1995,monaghan_SPHTensileInstability_2000,lyu_RemovingNumericalInstability_2021}.
This reduces numerical stability and accuracy.
In this regard, a wide stencil is required for the stable calculation of derivative values; however, this results in a large computational cost compared with mesh-based methods.

One of the strategies to maintain the uniformity of the particle distribution is particle shifting.
As pioneering research, Xu et al. \cite{xu_AccuracyStabilityIncompressible_2009} proposed particle shifting based on the diffusion equation in the context of the projection-based incompressible SPH method \cite{shao_IncompressibleSPHMethod_2003}.
The approach was generalized and applied to free surface flows \cite{lind_IncompressibleSmoothedParticle_2012}.
The core idea of these methods is to correct particle positions using a particle shifting vector, which is given by a diffusion flux in Fick's laws for the concentration of clustered particles.
However, the researchers indicated that the original formulation does not satisfy conservation features for momentum, angular momentum, or volume because particles are shifted regardless of any physical consistency in fluid mechanics \cite{lind_IncompressibleSmoothedParticle_2012,lyu_FurtherEnhancementParticle_2022}.
Several extended approaches have been proposed (e.g., \cite{lyu_FurtherEnhancementParticle_2022,gao_NewParticleShifting_2023,morikawa_CorrectedALEISPHNovel_2023}).
To essentially avoid the inconsistency, Oger et al. \cite{oger_SPHAccuracyImprovement_2016} applied an arbitrary Lagrangian--Eulerian (ALE) formulation \cite{hirt_ArbitraryLagrangianEulerianComputing_1974} for particle methods with particle shifting.
Several variations based on the ALE formulation have been proposed \cite{matsunaga_ImprovedTreatmentWall_2020,oger_SPHAccuracyImprovement_2016,morikawa_CorrectedALEISPHNovel_2023,hu_ALEParticleMethod_2017,antuono_DALESPHModelArbitrary_2021,matsunaga_StabilizedLSMPSMethod_2022,wang_CompactMovingParticle_2023,rastelli_ArbitrarilyLagrangianEulerian_2023}.

The ALE formulation also has the potential to reduce the bottleneck for load imbalance in parallel computation by controlling particle positions adequately.
Despite this, even when efficient parallelization techniques \cite{murotani_PerformanceImprovementsDifferential_2015,sodersten_BucketbasedMultigridPreconditioner_2019,kondo_ScalablePhysicallyConsistent_2023} are used, it is still challenging to establish a universal (i.e., problem-independent) method that achieves a good balance for calculation and node-to-node communication because neighboring information about the particle arrangement should be reassigned according to particle motions.

The mesh-constrained discrete point (MCD) method was designed to mitigate aforementioned problems \cite{matsuda_ParticlebasedMethodUsing_2022}.
Although the MCD method is a meshless method that uses MLS approximation, the arrangement of particles (or DPs) is specialized so that their positions are constrained in a background mesh to obtain a closely uniform distribution.
This achieves a reasonable approximation for spatial derivatives with compact stencils without encountering any ill-posed condition and leads to good performance in terms of computational efficiency.
Unlike existing hybrid methods that use particle and mesh systems, such as the material point method (\cite{sulsky_ParticleMethodHistorydependent_1994,li_SloshingImpactSimulation_2014,song_NonpenetrationFEMMPMContact_2020,li_ImmersedFiniteElement_2022,huang_LargedeformationSimulationsRoot_2024}) and particle-grid hybrid method (\cite{zhang_HyPAMHybridContinuumparticle_2009,matsunaga_HybridGridparticleMethod_2015,liu_ConservativeFiniteVolumeparticle_2019,xu_ThreedimensionalISPHFVMCoupling_2023,yao_FreeSurfaceTension_2024}), the MCD method only uses the background mesh to determine the DP positions, and does not need any interpolation between particles and mesh systems, which appears in other hybrid methods.
At present, the MCD method is limited to unsteady Stokes flows with stationary boundaries, and a novel formulation needs to be applied for moving boundary flow problems.

In this study, a novel meshless method is proposed based on the MCD method for incompressible flows with moving boundaries.
To ensure the mesh constraint of each DP in moving boundary problems, a novel updating algorithm for the DP arrangement is developed so that the position of DPs is not only rearranged but the DPs are also reassigned the role of being on the boundary or not.
Because each DP is prohibited from moving outside the background mesh element, capturing a moving interface using the same DPs is not possible.
This problem is solved by changing the DP roles (domain/boundary/void) according to information about the relative positions between background mesh elements and moving boundaries, and controlling the DP arrangement along interface motions.
The proposed method achieved reasonable results in numerical experiments for well-known moving boundary problems.

This paper consists of the following sections.
In Section \ref{MCD_method}, a DP arrangement algorithm is presented for moving boundaries based on the MCD method.
In Section \ref{flow_solver}, a fluid solver based on a pressure projection method is described for two-dimensional (2D) Navier--Stokes equations with moving boundaries in the ALE formulation.
In Section \ref{numerical_test}, 2D numerical examples are considered to validate the proposed method.
Additionally, the effects of the gap resolution between moving boundaries on the numerical stability and accuracy of the proposed method are discussed.
In Section \ref{conclusion}, the conclusion and remarks are presented.

\section{MCD method} \label{MCD_method}
\subsection{Overview}
In the MCD method, the DPs for solution unknowns are dispersedly arranged in the fluid domain and (moving) boundary that allows to represent arbitrary boundary shapes accurately.
A unique feature of the MCD method is to introduce a background mesh system to constrain the DP arrangement, where each DP is assigned to a mesh element and prohibited from moving over corresponding mesh edges.
A mechanical-based relaxation algorithm is applied to maintain smooth distribution of DPs and strict restriction algorithm is applied for satisfying the mesh constraint.
Although a main idea has been established in \cite{matsuda_ParticlebasedMethodUsing_2022}, in this study, little modification for the DP masking and arrangement is introduced to simplify the algorithm.
To manage DPs with moving boundaries, a DP relocation algorithm is applied in combination with the mesh-constrained approach.
In this study, an arbitrary shape is given by a discrete signed distance function (SDF), that simplifies to treat a non-deformable moving body and to combine with the DP masking and arrangement algorithms.

\subsection{DP masking algorithm} \label{mask_DP}
In the MCD method, to identify mesh elements and DPs for computation with moving boundaries, ``mask'' is introduced.
In \cite{matsuda_ParticlebasedMethodUsing_2022}, temporal points (TPs) were introduced on the centroids of each background mesh element to determine the mask according to the SDF between the TP and boundary.
In the present study, an idea is developed to assign the mask based on the positional relationship between mesh vertices and moving boundaries, rather than using TPs.
A single DP is set in each mesh element, that is, the same mask is assigned to both the DP and mesh.
Let $n_{i,\,j}^{\rm mesh}$ denote the number of vertices, at where the SDF is greater than or equal to zero, for a mesh ($i,~ j$).
In a 2D uniform Cartesian mesh system with mesh length $l_{0}$, $n_{i,\,j}^{\rm mesh}$ can take values from 0 to 4 because the mesh ($i,~ j$) has four vertices: ($i,~ j$), ($i+1,~ j$), ($i,~ j+1$), and ($i+1,~ j+1$).
The mask at ($i,j$) is defined as
\begin{equation}
  {\rm mask}_{i, \, j} = \left\{ \begin{array}{ll}
    0,             & \text{for}~ n_{i,\,j}^{\rm mesh} = 0,~ 1, \\
    1,~ 2,~ ...,   & \text{for}~ n_{i,\,j}^{\rm mesh} = 2,~ 3, \\
    -1, & \text{for}~ n_{i,\,j}^{\rm mesh} = 4,
  \end{array} \right.
\end{equation}
where the value indicates that ${\rm mask}_{i,\,j} \geq 1$ for (moving) boundaries; ${\rm mask}_{i,\,j} = -1$ for the inner domain; and ${\rm mask}_{i,\,j} = 0$ for the void.
Note that different values for ${\rm mask}_{i,\,j} \geq 1$ can be used to identify multiple boundaries.

\subsection{DP arrangement algorithm}
In the mesh-constrained approach, an arrangement of DPs is generated given a constraint: each DP does not exceed the mesh edges.
The fundamental flow for the DP arrangement algorithm is as follows:
\begin{itemize}
  \item[(i)]   Set an initial DP arrangement at the centroid $\mathbf{x}_{\rm CM}$ of each background mesh element ($\mathbf{x}^{0} = \mathbf{x}_{\rm CM}$).
  \item[(ii)]  Move the DPs uniformly in space by solving a mechanical equilibrium problem ($\mathbf{x}^{k} \rightarrow \mathbf{x}^{*}$).
  \item[(iii)] Relocate the DPs onto (moving) boundaries based on the SDF ($\mathbf{x}^{*} \rightarrow \mathbf{x}^{**}$).
  \item[(iv)]  Correct the DPs that exceed the mesh edges ($\mathbf{x}^{**} \rightarrow \mathbf{x}^{k+1}$).
  \item[(v)]   Increment $k \rightarrow k+1$ and iterate procedures (ii)--(iv) until convergence.
\end{itemize}
Further details of each process are described in the following.

\begin{figure}[H] 
  \begin{center}
    \includegraphics[width=\linewidth]{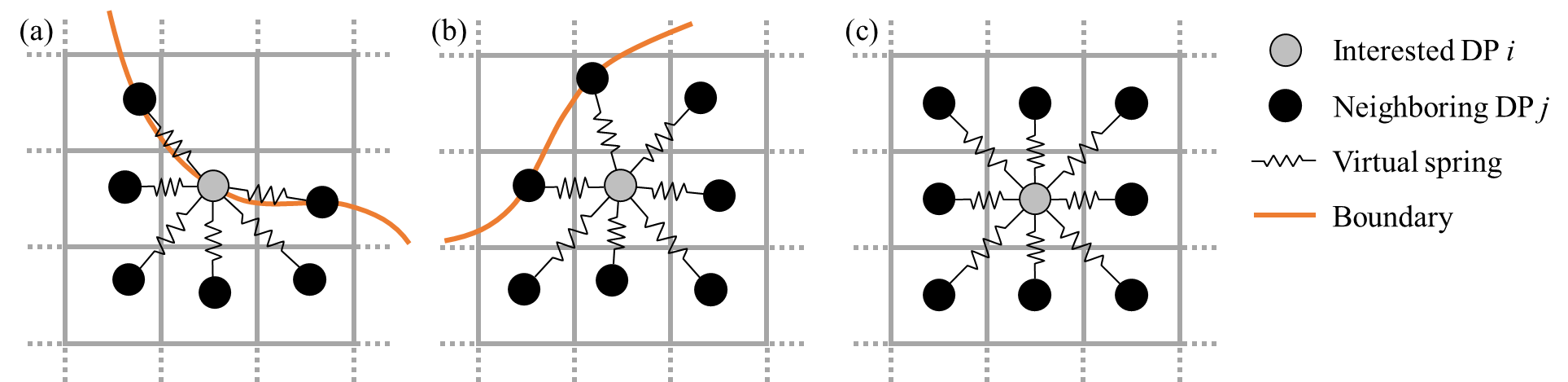}
    \caption{%
      Examples of interactions between the $i$-th DP and its neighboring $j$-th DPs by considering virtual connected springs in cases in which the $i$-th DP is located on $\Gamma$ (a), in $\Omega_{\rm I}$ and near $\Gamma$ (b), and in $\Omega_{\rm I}$ and far from $\Gamma$ (c).
    }
    \label{fig_DP_arrangement_spring}
  \end{center}
\end{figure}

Regarding procedure (ii), a non-inertial equation of motion is solved:
\begin{equation}
  \gamma \mathbf{v}_i + \mathbf{F}_{i} = 0,
  \label{eq_gov_eq_mcd}
\end{equation}
where $\mathbf{v}_i = {\rm d} \mathbf{x}_i / {\rm d} t$ denotes the velocity of the $i$-th DP, $\gamma$ denotes the coefficient of the damping force which is linearly dependent on $\mathbf{v}_i$, and $\mathbf{F}_{i}$ denotes the resultant force of interactive forces $\mathbf{F}_{ij}$ between neighboring $j$-th DPs.
Let $D^{\rm DP}_{i} = D^{\rm DP} (\mathbf{x}_{i})$ denote the compact support domain, that is, a $3 \times 3$ local mesh system centered on $\mathbf{x}_{i}$.
Now, the sets of neighboring DPs, $\Lambda_{i}^{\rm I}$ and $\Lambda_{i}^{\Gamma}$, are defined as
\begin{equation}
  \Lambda_{i}^{\rm I} = \left\{ j \in [1,n_{\Omega}] \mid j \neq i, \, \mathbf{x}_j \in D_i^{\rm DP}, \, \mathbf{x}_j \in \Omega_{\rm I} \right\},
  \label{eq_set_of_neighboring_DPs_I}
\end{equation}
\begin{equation}
  \Lambda_{i}^{\Gamma} = \left\{ j \in [1,n_{\Omega}] \mid j \neq i, \, \mathbf{x}_j \in D_i^{\rm DP}, \, \mathbf{x}_j \in \Gamma \right\},
  \label{eq_set_of_neighboring_DPs_Gamma}
\end{equation}
where $n_{\Omega}$ denotes the number of DPs in the analysis domain $\Omega$, and $\Omega_{\mathrm I} = \Omega \setminus \Gamma$ denotes the inner domain excluding the boundary $\Gamma$.
The resultant force $\mathbf{F}_{i}$ can be written as
\begin{equation}
  \mathbf{F}_{i} = \sum_{ j \in \Lambda_{i}^{\rm I} \cup \Lambda_{i}^{\Gamma} }{ \mathbf{F}_{ij} }.
  \label{eq_resultant_force}
\end{equation}
Note that, in \cite{matsuda_ParticlebasedMethodUsing_2022} the calculations of forces acting on each DP in the domain and on the boundary in the different procedures were addressed.
The DP in $\Omega_{\rm I}$ receives interaction forces by surrounding DPs in $\Omega_{\rm I} \cup \Gamma$, and the DP on $\Gamma$ receives interaction forces by only neighboring DPs on $\Gamma$.
In this study, both the DPs in $\Omega_{\rm I}$ and $\Gamma$ are simultaneously updated using Eq. (\ref{eq_resultant_force}), rather than the aforementioned step-by-step procedure, to improve the uniformity of the DP arrangement.
The interaction force is calculated as $\mathbf{F}_{ij} = F_{0} \hat{F}_{ij} \tilde{\mathbf{x}}_{ij}$, where $F_{0}$ denotes the constant force coefficient, $\tilde{\mathbf{x}}_{ij} = \mathbf{x}_{ij} / || \mathbf{x}_{ij} ||$ denotes the unit directional vector of $\mathbf{x}_{ij} = \mathbf{x}_{j} - \mathbf{x}_{i}$, and $\hat{F}_{ij} = G (r')$ denotes the dimensionless force function.
In \cite{matsuda_ParticlebasedMethodUsing_2022}, $G (r')$ was considered only for the repulsive force, whereas in the present study, a virtual spring is considered between DPs and the restoring force is used to enhance the homogenization of the DP arrangement (Figure \ref{fig_DP_arrangement_spring}):
\begin{equation}
  G (r') = G \left( ||\mathbf{x}_{ij}|| / l_{0}^{\rm{SPG}} \right) = 1 - r'.
  \label{eq_dimensionless_force_function}
\end{equation}
where $l_{0}^{\rm{SPG}}$ denotes the natural length of the virtual spring,
\begin{equation}
  l_{0}^{\rm{SPG}} = \left\{
  \begin{array}{cl}
    \sqrt{2} l_{0}, & {\text{if the neighboring DPs are on the diagonal-side grids}},  \\
    l_{0}, & \text{otherwise}.
  \end{array} \right.
  \label{eq_natural_length_virtual_spring}
\end{equation}
This virtual spring is expected to align DPs that are far from boundaries as a regular-grid arrangement (Figure \ref{fig_DP_arrangement_spring}-c) and suppress the condition number in MLS approximations described later.

The application of the explicit Euler method with the time interval $\Delta \tau$ to Eq. (\ref{eq_gov_eq_mcd}) yields the dimensionless first intermediate position $\hat{\mathbf{x}}_{i}^{*} = \mathbf{x}_{i}^{*} / l_{0}$:
\begin{equation}
  \hat{\mathbf{x}}_{i}^{*} = \hat{\mathbf{x}}_{i}^{k} - \kappa \hat{\mathbf{F}}_{i}^{k},
  \label{eq_1st_intermed_position_x*}
\end{equation}
where $\kappa = \Delta \tau F_{0} / \gamma l_{0}$ denotes the dimensionless coefficient.
When $\kappa$ is set to a certain value, the proposed algorithm exhibits similar convergence behavior regardless of the mesh length \cite{matsuda_ParticlebasedMethodUsing_2022}.
In this study, $\kappa$ is set to $\kappa = 1 /5$.

In procedure (iii), the DP arrangement is made to follow the boundary shape using the SDF for the boundary surface $\Gamma$ as follows:
\begin{equation}
  \mathbf{x}^{**}_{i} = 
  \left\{ \begin{array}{cl}
    \mathbf{x}^{*}_{i}, & \left( i \in \Lambda^{\rm I}_{i} \right), \\[2mm]
    \mathbf{x}^{*}_{i} + \psi_{i} \hat{\mathbf{d}}_i, & \left( i \in \Lambda^{\Gamma}_{i} \right),
  \end{array} \right.
  \label{eq_2nd_intermed_position_x**}
\end{equation}
where $\psi_i$ denotes the SDF value at $\mathbf{x}_{i}^{*}$ and $\hat{\mathbf{d}}_{i} = - \nabla \psi_{i}/||\nabla \psi_{i}||$ denotes the unit direction vector from $\mathbf{x}_{i}^{*}$ to the nearest boundary $\Gamma$.
These values are evaluated using MLS reconstruction at $\mathbf{x}^{*}_i$.

\begin{figure}[H] 
  \begin{center}
    \includegraphics[width=.7\linewidth]{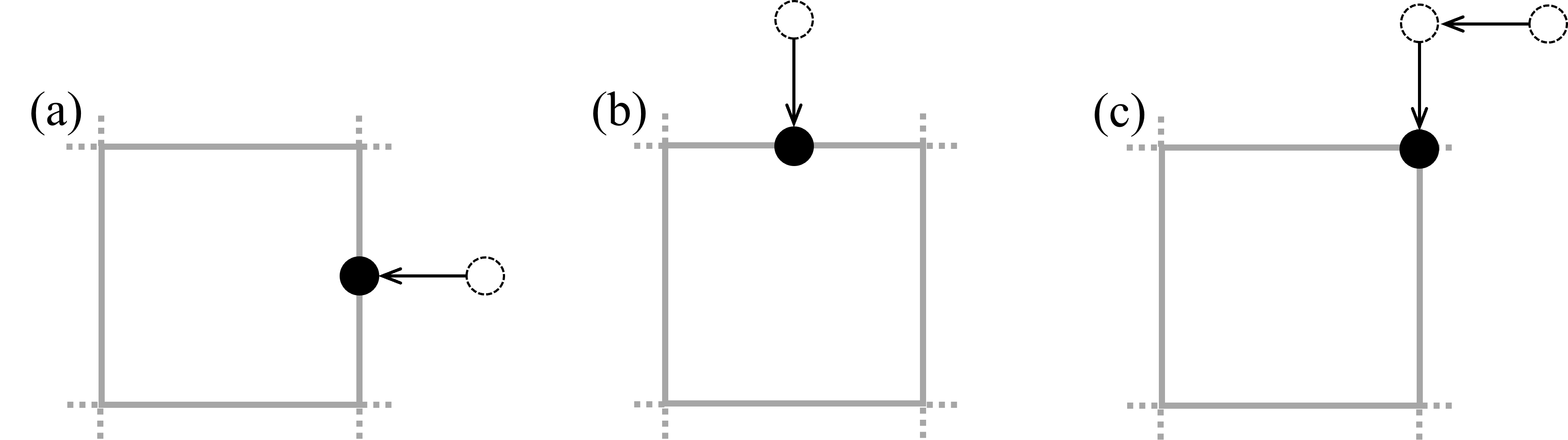}
    \caption{%
      Examples of the mesh constraint procedure for a DP that exceeds the element of background mesh corresponding to the DP.
      (a) When the DP (dashed line) is outside of the edges of the mesh element for the $x$-axis, its $x$-coordinate is replaced by that of the edges of the mesh element (black dot).
      (b) Same as (a), but for the $y$-axis.
      (c) When the DP (dashed circle) is outside of both edges of the mesh element for both $x$ and $y$-axes, its $x$ and $y$-coordinates are replaced by those of the edges of the mesh element, which relocates the DP to the nearby vertex of the mesh element (black dot).
    }
    \label{fig_mesh_constraint}
  \end{center}
\end{figure}

In procedure (iv), the mesh constraint is applied for DPs that exceed each background mesh element, which are relocated to the nearest edge or vertex of the background mesh element (Figure \ref{fig_mesh_constraint}).

\subsection{Treatment of moving boundaries}
\begin{figure}[H] 
  \begin{center}
    \includegraphics[width=\linewidth]{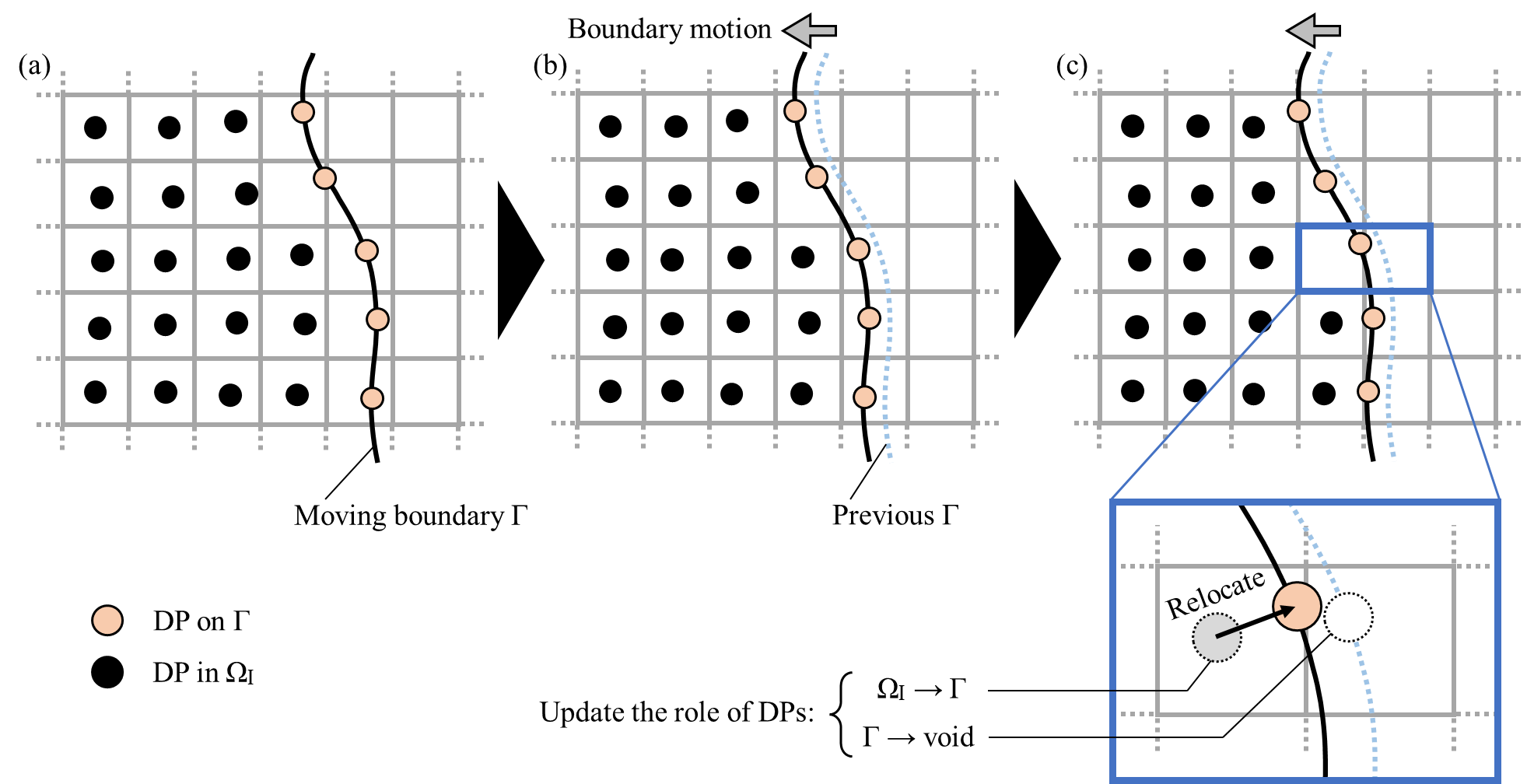}
    \caption{%
      Example of the mesh-constrained approach for a moving boundary.
      (a) DP arrangement around a moving boundary $\Gamma$ that moves toward the left.
      (b) Unless the relationship between the background mesh elements and moving boundary changes, the mask of each DP is maintained.
      (c) If a boundary moves beyond a mesh vertex, the mask and position of the DP is updated according to the new arrangement.
    }
    \label{fig_moving_boundary_treatment_MCD}
  \end{center}
\end{figure}

Compared with fully Lagrangian particle methods, the main issue in applying the proposed MCD method to moving boundary problems is how to manage the moving boundary using the mesh-constrained approach.
Figure \ref{fig_moving_boundary_treatment_MCD} shows a conceptual illustration of the proposed approach to update the DP arrangement for the moving boundary.
The DPs are masked according to the positional relationship between each element of the background mesh and the moving boundary.
A moving boundary that moves toward the left (Figure \ref{fig_moving_boundary_treatment_MCD}-a) is considered.
If the relationships between each background mesh element and moving boundary are not changed, the mask of each DP is maintained (Figure \ref{fig_moving_boundary_treatment_MCD}-b).
When the boundary moves across the mesh vertices, the masks of DPs are updated according to a new arrangement (Figure \ref{fig_moving_boundary_treatment_MCD}-c).
As in the enlarged part of the figure, the mask of the DP that was on $\Gamma$ at the previous time changed from ``boundary'' to ``void'' and the DP was excluded from the calculation, whereas the mask of the DP that was in $\Omega_{\rm I}$ at the previous time changed from ``void'' to ``boundary'' and the position was relocated onto $\Gamma$.
A similar procedure was adopted for the DPs located on the opposite side of the direction of travel, that is, the mask of the DPs changed from ``boundary'' to ``void'' or ``inner'' to ``boundary.''

In this study, to reduce the computational cost, the mask and position are only redefined for DPs located near the boundary whose distance from the boundary is less than $5l_0$.

\subsection{Representation of arbitrary boundary shapes with a SDF} \label{sdf}
\begin{figure}[H] 
  \begin{center}
    \includegraphics[width=.6\linewidth]{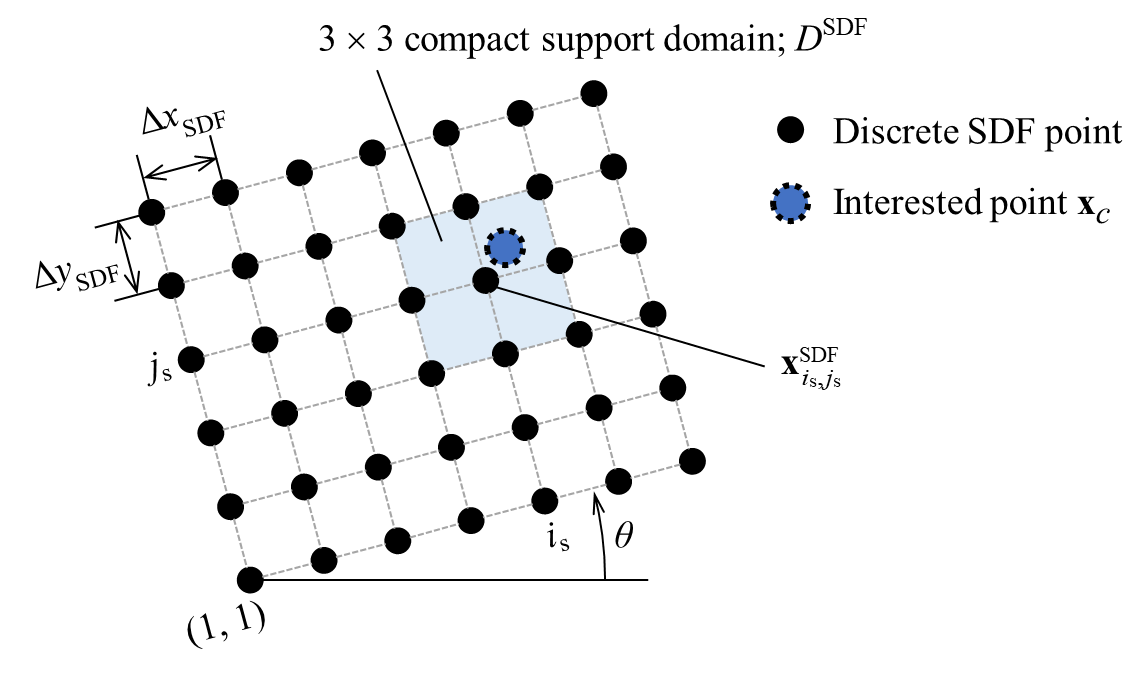}
    \caption{%
      Interpolation of the SDF at an interested point $\mathbf{x}_{c}$ in a compact support domain $D^{\rm SDF}$ for the nearest points $\mathbf{x}_{i_{\rm s}, \, j_{\rm s}}^{\rm SDF}$.
    }
    \label{fig_SDF_nearest_point}
  \end{center}
\end{figure}

The arbitrary boundary is represented by the SDF $\psi = \psi ({\mathbf{x}})$, which is $\psi > 0$ for the inner domain ($\mathbf{x} \in \Omega_{\rm I}$), $\psi < 0$ outside the domain ($\mathbf{x} \notin \Omega$), and $\psi = 0$ on the boundary ($\mathbf{x} \in \Gamma$).
The SDF is discretely given by $\psi_{i, \, j}$ on a uniform Cartesian mesh system $\mathbf{x}^{\rm SDF}_{i, \, j} = (x_{i}^{\rm SDF},~ y_{j}^{\rm SDF})$ for $i \in [1, N_{x}^{\rm SDF}],~ j \in [1, N_{y}^{\rm SDF}]$.
It is easy to evaluate the SDF $\psi_{c} = \psi (\mathbf{x}_{c})$ and spatial derivatives $\nabla \psi_{c}$ using the MLS interpolation described in Section \ref{spatial_discretization}.
Let $D^{\rm SDF}$ denote a $3\times3$ compact support domain for interpolating $\psi_{c}$ from the discrete SDF cloud, where the center of the domain $\mathbf{x}_{i_{\rm s}, \, j_{\rm s}}^{\rm SDF}$ is defined as the nearest point of $\mathbf{x}_{c}$.
Because non-deformed boundaries are assumed in this study, a local coordinate system with posture angle $\theta$ for a body is adopted (Figure \ref{fig_SDF_nearest_point}).

In the case of multiple boundaries, each corresponding SDF is introduced, and the nearest SDF, which has the smallest magnitude value, is adopted.

\section{MCD method for the incompressible Navier--Stokes equations} \label{flow_solver}
\subsection{Governing equations}
In this study, the arrangement of DPs is updated with time following moving boundaries. Thus, the Navier--Stokes equation based on the arbitrary ALE formulation is used:
\begin{equation}
  \nabla \cdot \mathbf{v} = 0,
  \label{eq_continuity}
\end{equation}
\begin{equation}
  \frac{\partial \mathbf{v}}{\partial t} + \left( \mathbf{v} - \mathbf{w} \right) \cdot \nabla \mathbf{v} = - \nabla P + \nu \nabla^{2} \mathbf{v}.
  \label{eq_ale_ns}
\end{equation}
Here, $\mathbf{v}$ is the velocity vector, $P$ is the pressure normalized by the fluid density $\rho$, and $\nu=\mu/\rho$ is the kinematic viscosity, where $\mu$ is the dynamics viscosity.
Additionally, $t$ denotes time, $\nabla = \partial / \partial \mathbf{x}$ denotes the spatial differential operator, and $\mathbf{w}$ denotes the velocity of DPs (ALE velocity).

\subsection{MLS approximation for spatial derivatives} \label{spatial_discretization}
Now the evaluation of the discrete quantity $\phi_{c}$ at an arbitrary position $\mathbf{x}_{c}$ is considered.
In the MCD method, DPs can be distributed in arbitrary positions around boundary shapes under the mesh constraint condition.
Hence, in this study, the generalized quadratic MLS formulation described in \cite{matsuda_ParticlebasedMethodUsing_2022} is used for evaluating a discretized quantity and its spatial differentiation.
The MCD method can use the structure information for adjacent DPs using the background mesh system and execute stencil calculations efficiently, which is a different feature from other particle-based methods that have no DP adjacent information.
In this study, 2D systems are calculated using a compact support domain $D_c$ of $3 \times 3$ background mesh elements for arbitrary point $\mathbf{x}_{c}$.

An approximation of the arbitrary quantity $\phi$ is defined by a polynomial function: 
\begin{equation}
  \phi^{h}({\bf X}) = \phi_{c} + \mathbf{p}({\bf X}) \cdot {\bm \Phi}_{c},
  \label{eq_intrp_chi_1}
\end{equation}
where ${\bf X}=({\bf x}-{\bf x}_{c})/r_{s}$ is the relative position with respect to the position ${\bf x}_{c}$ and the scaling parameter $r_{s}$ (set to $r_{s} = l_{0}$ in this study), $\mathbf{p} (\mathbf{X})$ is the polynomial basis vector, and ${\bm \Phi}_{c}$ is the vector of polynomial coefficient (or normalized differential coefficient at ${\bf x}_{c}$).
In the case of the quadratic polynomial in 2D, where $\mathbf{X} = (X,~ Y)$, $\mathbf{p} (\mathbf{X})$, and ${\bm \Phi}_{c}$ are given by
\begin{equation}
  \mathbf{p} (\mathbf{X}) 
  = \left( X, ~ Y, ~ \frac{1}{2}X^2, ~ XY, ~ \frac{1}{2}Y^2 \right),
  \label{eq_p_ic}
\end{equation}
\begin{equation}
  {\bm \Phi}_{c} = 
  \left(
    \left. \frac{\partial \phi}{\partial X} \right|_{c},
    ~ \left. \frac{\partial \phi}{\partial Y} \right|_{c},
    ~ \left. \frac{\partial^2 \phi}{\partial X^2} \right|_{c},
    ~ \left. \frac{\partial^2 \phi}{\partial X \partial Y} \right|_{c},
    ~ \left. \frac{\partial^2 \phi}{\partial Y^2} \right|_{c}
  \right).
  \label{eq_spatial_derivative_phi_i}
\end{equation}
Let $\Gamma_{\rm D}$ denote the Dirichlet boundary, $\Gamma_{\rm N}$ denote the Neumann boundary, and $\Omega_{\rm I} = \Omega \setminus \Gamma$ denote the inner domain.
The objective function for the MLS reconstruction can be written as
\begin{equation}
  J = \frac{1}{2} \!\!\! \sum_{
      {\fontsize{7pt}{0mm}\selectfont
        \begin{array}{l}
          j \! \in \! \Lambda_c
        \end{array}}
      }{\!\!\!
      w_{j} \left( \mathbf{p}_{j} \cdot {\bm{\Phi}}_{c} +\phi_c -\phi_j \right)^2}
    + \frac{1}{2} \!\!\! \sum_{
      {\fontsize{7pt}{0mm}\selectfont
        \begin{array}{l}
          j \! \in \! \Lambda^{\rm N}_c
        \end{array}}
      }{\!\!\!
      w_{j} \left\{ r_{\rm s} \left( \frac{1}{r_{\rm s}} \mathbf{p}^{\rm N}_{j} \cdot {\bm{\Phi}}_{c} -f_{j} \right) \right\}^{2}}
    + \chi_{c} \lambda_{c} (\phi_{c} - g),
  \label{eq_obj_func}
\end{equation}
where $\phi_{j}$ is the discrete value of $\phi$ for the $j$-th DP, $\mathbf{p}_{j} = \mathbf{p}({\bf X}_{j})$ for ${\bf X}_{j}=({\bf x}_{j}-{\bf x}_{c})/r_{s}$, $\mathbf{p}^{\rm N}_{j} = \partial_{\rm N}\mathbf{p}({\bf X}_{j})$ is the polynomial basis vector with respect to the normal derivative $\partial_{\rm N} = {\bf n}\cdot\nabla_{\bf X}$, and
\begin{equation}
  \Lambda_c = \left\{ 
    i \in [1,n_{\Omega}] \mid \mathbf{x}_i \in D_c, \mathbf{x}_i \in \Omega_{\rm I} \cup \Gamma_{\rm D} \right\},~~
  \Lambda^{\rm N}_c = \left\{ 
    i \in [1,n_{\Omega}] \mid \mathbf{x}_i \in D_c, \mathbf{x}_i \in \Gamma_{\rm N} \right\}.
  \label{eq_set}
\end{equation}
In this study, $w_{j} = w \left( r_{e};~ ||\mathbf{x}_{j} - \mathbf{x}_{c}|| \right)$ denotes the weight, in which function $w$ is defined as
\begin{equation}
  w_{j}
  = \left\{ \begin{array}{cl} \displaystyle
    \cos^{2}{\left( \frac{\pi ||\mathbf{x}_{j} - \mathbf{x}_{c}||}{2r_{e}} \right)},   & {\rm{for}}~ ||\mathbf{x}_{j} - \mathbf{x}_{c}|| \leq r_{e}, \\[12pt]
    0,                                                                                 & {\rm{otherwise}},
  \end{array} \right.
\end{equation}
where $r_{e}$ denotes the influence radius for the MLS reconstruction and is set to $r_{e} = 2.5 l_{0}$.
A normal equation of the MLS can be obtained, and the physical quantity and its spatial gradient can be interpolated by calculating the stationary condition; $\partial J / \partial \phi_{c} = 0$, $\partial J / \partial {\bm{\Phi}}_{c} = 0$, $\partial J / \partial \lambda_{c} = 0$.
Regarding Eq. (\ref{eq_obj_func}), the second term on the right-hand side was introduced in \cite{matsunaga_ImprovedTreatmentWall_2020} and reflects the Neumann boundary condition.
The third term on the right-hand side is the constraint condition determined by the Lagrangian multiplier method.
In this study, $\lambda_{c}$ is the Lagrangian multiplier and $\chi_{c}$ is the switching parameter that indicates whether the constraint of the Lagrangian multiplier is effective; the value is 0 or 1; that is, the proposed MLS approximation has two types of formulations: $\chi_{c} = 0$ or $\chi_{c} = 1$.
The $\chi_{c} = 0$ formulation is the MLS approximation not including values at the arbitrary stencil-centric position $\mathbf{x}_{c}$ (or $\phi_{c}$ is unknown).
The formulation of $\chi_{c} = 1$ is the MLS fitting including values on stencil-centric position $\mathbf{x}_{c}$ (i.e., $g = \phi_{c}$).
Readers can see more details in \cite{matsuda_ParticlebasedMethodUsing_2022}.
In this study, only the condition $\chi_{c}=1$, where a stencil-center is also equivalent to the $i$-th DP (i.e., ${\bf x}_{c} = {\bf x}_{i}$ and $g = \phi_{c} = \phi_{i}$), is used.

\subsection{Velocity--pressure coupling} \label{vel_pres_cpl}
To solve the incompressible Navier--Stokes equation (Eqs. (\ref{eq_continuity}) and (\ref{eq_ale_ns})), a fractional step method (pressure projection method) \cite{chorin_NumericalSolutionNavierStokes_1968,matsunaga_StabilizedLSMPSMethod_2022} is applied.

First, the displacement of DPs $\Delta \mathbf{x}^{k}$ is calculated:
\begin{equation}
  \Delta \mathbf{x}^{k} = \mathbf{x}^{k+1} - \mathbf{x}^{k}.
  \label{eq_dx}
\end{equation}
Note that the position of DPs is not updated yet in this sequence.
Then, the velocity is interpolated at the updated position $\mathbf{x}^{k+1}$ based on the Taylor series expansion at the center of $\mathbf{x}^{k}$ so that the ALE velocity term is considered \cite{matsunaga_StabilizedLSMPSMethod_2022}, that is, $\tilde{\mathbf{u}}^{k} = \mathbf{u}^{k} (\mathbf{x}^{k+1})$:
\begin{equation}
  \tilde{\mathbf{u}}^{k} = \mathbf{u}^{k} + \sum_{m=1}^{p}{\frac{1}{m!}\left( \Delta \mathbf{x}^{k} \cdot \nabla \right)^{m} \mathbf{u}^{k}}.
  \label{eq_ale_vel}
\end{equation}
In this study, the series is expanded to the second-order ($p = 2$), which is equivalent to interpolation using a quadratic MLS reconstruction for $\chi_{c}=1$ at ${\bf x}^k$.
Note that, in the framework of the proposed method, the DP not only moves along the moving boundary but also changes its mask.
The case is considered that the DP on the boundary at $k$-th step becomes the one on the fluid domain at the $k+1$-th step.
In this case, the physical information (e.g., pressure or velocity) of the DP after its mask is changed is missing.
To uniformly obtain $\tilde{\mathbf{u}}^{k}$ of the DP cloud in $\Omega_{\rm I}$, including the DP that is missing information, the previous (or $k$-th) mask, position, and velocity are used as the stencil calculation for the MLS interpolation (\ref{appendix_A}).

After $\tilde{\mathbf{u}}$ is calculated, the position of the DPs is updated to $\mathbf{x}^{k} \rightarrow \mathbf{x}^{k+1}$ and the next timestep of velocity $\mathbf{u}^{k+1}$ and pressure $P^{k+1}$ is calculated using the same method as the well-known fractional step method \cite{chorin_NumericalSolutionNavierStokes_1968}:
\begin{equation}
  \mathbf{u}^{*} = \mathbf{F}^{k}_{\rm{MOC}} + \nu \Delta t \nabla^{2} \tilde{\mathbf{u}}^{k},
  \label{eq_intermed_vel}
\end{equation}
\begin{equation}
  \nabla^{2} P^{k+1} = \frac{1}{\Delta t} \nabla \cdot \mathbf{u}^{*},
  \label{eq_pressure_poisson}
\end{equation}
\begin{equation}
  \mathbf{u}^{k+1} = \mathbf{u}^{*} - \Delta t \nabla P^{k+1},
  \label{eq_projection}
\end{equation}
where $\mathbf{u}^{*}$ is the intermediate velocity.
$\mathbf{F}^{k}_{\rm{MOC}} = \mathbf{F}^{k}_{\rm{MOC}}[\mathbf{x}^{k+1} - \tilde{\mathbf{u}}^{k} \Delta t;~ \tilde{\mathbf{u}}^{k}]$ is related to the advection term approximated by a first-order temporal discretization using the method of characteristics, that is, $\mathbf{F}^{k}_{\rm{MOC}}$ is the upwind value of the velocity at $\mathbf{x}^{k+1}-\tilde{\mathbf{u}}^{k} \Delta t$ evaluated using a similar approach to that in Eq. (\ref{eq_ale_vel}).
In this study, the linear systems for the pressure Poisson equation (Eq. (\ref{eq_pressure_poisson})) are calculated using the Bi-CGSTAB method.
The algebraic equation of the pressure Poisson equation is obtained using the MLS approximation with Neumann boundary condition on the moving boundary, which is given based on the formulation \cite{udaykumar_SharpInterfaceCartesian_2001},
\begin{equation}
  \frac{\partial P}{\partial n} = - \frac{{\rm d} \mathbf{V}^{k+1}}{{\rm d} t} \cdot \mathbf{n}^{k+1},
  \label{eq_neumann_moving_boundary}
\end{equation}
where $\mathbf{n}$ denotes the outward unit normal vector and $\mathbf{V}$ denotes the velocity of the moving boundary.

In this study, the moving boundary of a rigid body is considered; thus, the velocity of the moving body is given as the summation of translational and rotational velocities:
\begin{align}
  \mathbf{V} &= \mathbf{V}_{\rm trans} + \mathbf{V}_{\rm rot} \\
    &= \mathbf{V}_{\rm trans} + \bm{\omega} \times \mathbf{r},
  \label{eq_velocity_moving_boundary}
\end{align}
where $\bm{\omega}$ denotes the angular velocity and $\mathbf{r} ~(= \mathbf{x}_{\Gamma} - \mathbf{x}_{\rm c})$ denotes the direction vector from the centroid.
The temporal derivative of $\mathbf{V}$ at $k+1$ is approximated by
\begin{align}
  \frac{{\rm d} \mathbf{V}^{k+1}}{{\rm d} t}
  &= \frac{{\rm d} \mathbf{V}_{\rm trans}^{k+1}}{{\rm d} t}
  + \frac{{\rm d} \bm{\omega}^{k+1}}{{\rm d} t} \times \mathbf{r}^{k+1}
  + \bm{\omega}^{k+1} \times \frac{{\rm d} \mathbf{r}^{k+1}}{{\rm d} t} \\
  &= \frac{{\rm d} \mathbf{V}_{\rm trans}^{k+1}}{{\rm d} t}
  + \frac{{\rm d} \bm{\omega}^{k+1}}{{\rm d} t} \times \mathbf{r}^{k+1}
  + \bm{\omega}^{k+1} \times \left( \frac{{\rm d} \mathbf{x}_{\Gamma}^{k+1}}{{\rm d} t} - \frac{{\rm d} \mathbf{x}_{\rm c}^{k+1}}{{\rm d} t} \right) \\
  &= \frac{{\rm d} \mathbf{V}_{\rm trans}^{k+1}}{{\rm d} t}
  + \frac{{\rm d} \bm{\omega}^{k+1}}{{\rm d} t} \times \mathbf{r}^{k+1}
  + \bm{\omega}^{k+1} \times \mathbf{V}_{\rm rot}^{k+1} \\
  &\approx %
  \frac{\mathbf{V}_{\rm trans}^{k+1} - \mathbf{V}_{\rm trans}^{k}}{\Delta t} + %
  \frac{\bm{\omega}^{k+1} - \bm{\omega}^{k}}{\Delta t} \times \mathbf{r}^{k+1}
  + \bm{\omega}^{k+1} \times \mathbf{V}_{\rm rot}^{k+1},
  \label{eq_dvdt_moving_boundary}
\end{align}
where, ${\rm d} \mathbf{x}_{\Gamma} / {\rm d} t = \mathbf{V}_{\rm trans} + \mathbf{V}_{\rm rot}$ and ${\rm d} \mathbf{x}_{c} / {\rm d} t = \mathbf{V}_{\rm trans}$.

\subsection{Evaluation of fluid forces acting on moving boundaries} \label{eval_fluid_forces_mov_bound}
\begin{figure}[H] 
  \begin{center}
    \includegraphics[width=.6\linewidth]{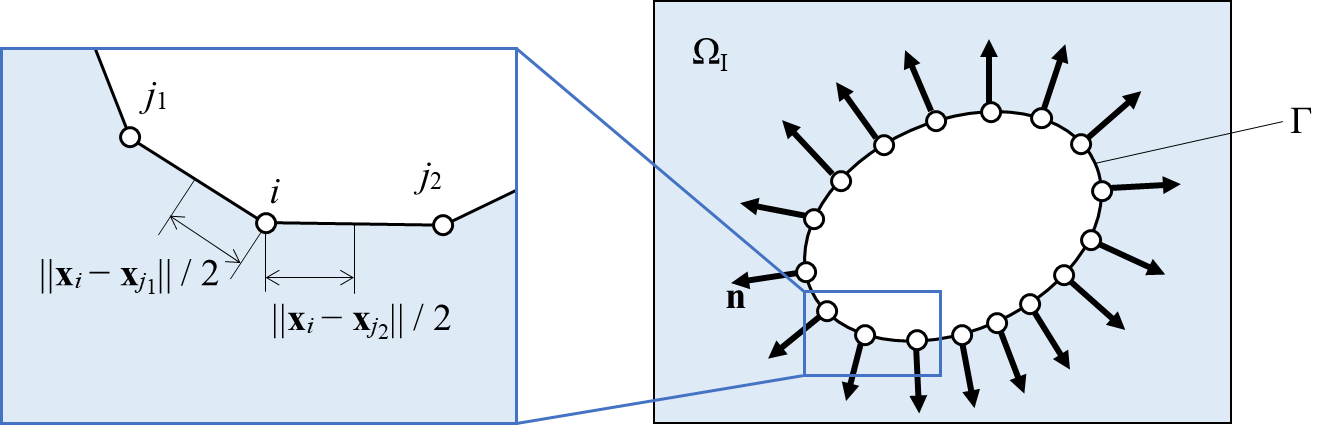}
    \caption{%
      Evaluation of the surface element of boundary $\Gamma$ in an approximation of fluid forces acting on a moving boundary.
    }
    \label{fig_area_element}
  \end{center}
\end{figure}

As shown in Figure \ref{fig_area_element}, fluid force $\mathbf{F}$ defined by a surface integral of a surface force vector $\mathbf{t}= (-P {\bf I} + \mu ( \nabla \mathbf{u} + \nabla \mathbf{u}^{\top} ) ) \cdot \mathbf{n}$ is approximated by
\begin{equation}
  \mathbf{F} = \int_{\Gamma} \mathbf{t} \, {\rm{d}} \Gamma
  \approx \sum_{ i \, \in \, \Lambda_{\rm{\Gamma}}}{\mathbf{t}_{i} \, \Delta\Gamma_{i}},
  \label{eq_discretized_fluid_force}
\end{equation}
where $\Delta\Gamma$ is the surface element of boundary $\Gamma$, which is calculated as the arithmetic mean distance from $\mathbf{x}_{i}$ to the neighboring points $\mathbf{x}_{j_1}$ and $\mathbf{x}_{j_2}$ as follows:
\begin{equation}
  \Delta\Gamma_{i} = \frac{1}{2} \left( || \mathbf{x}_{i} - \mathbf{x}_{j_{1}} || + || \mathbf{x}_{i} - \mathbf{x}_{j_{2}} || \right).
  \label{eq_surface_area_element}
\end{equation}

\begin{figure}[H] 
  \begin{center}
    \includegraphics[width=\linewidth]{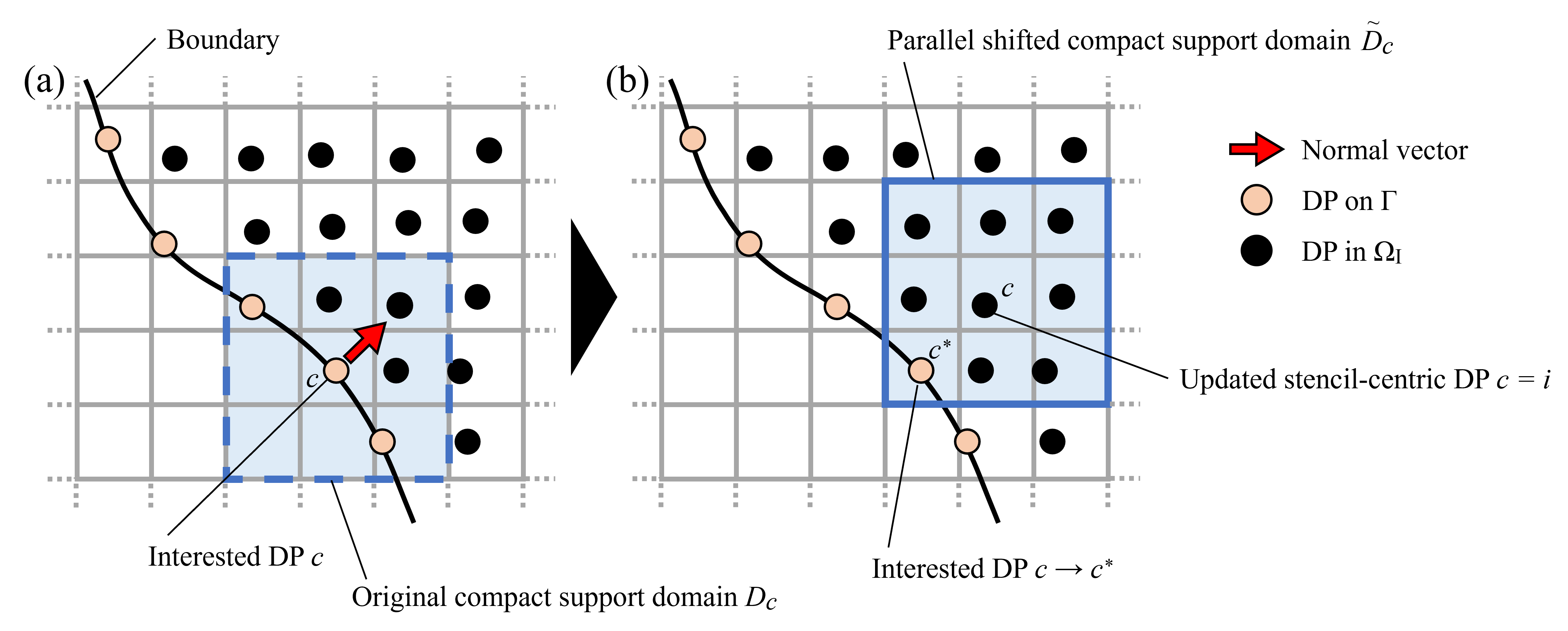}
    \caption{%
      Example of the translation of a $3 \times 3$ compact support domain for the MLS reconstruction on the boundary (to evaluate the surface force vector): (a) before translation; (b) after.
      Translation of the compact support domain $D_{c} \rightarrow \tilde{D}_{c}$ according to the signs of the $x$ and $y$-components of the boundary normal vector.
      For the $\chi_{c} = 1$ MLS reconstruction, a stencil-centric position is set to $\mathbf{x}_{c} = \mathbf{x}_{i}$.
    }
    \label{fig_stencil_shift}
  \end{center}
\end{figure}

To evaluate the surface force vector, the velocity gradients and pressure on the moving boundary $\Gamma$ are calculated using the MLS approximations.
In this regard, a slight modification is made to enable compact support in the MLS reconstruction.
If the symmetric $3 \times 3$ compact stencils are used for spatial discretization on the boundary, some mesh elements outside the analysis domain could be included in the compact support domain, which leads to a lack of stencils required for a stable MLS approximation (Figure \ref{fig_stencil_shift}-a).
To avoid a lack of stencils, translational shifts for the compact support domain are performed as $D_{c} \rightarrow \tilde{D}_{c}$ according to a boundary normal vector $\mathbf{n} = (n_{x},~ n_{y})$ on an interested $c$-th DP. 
In this study, $D_{c}$ shifts by one mesh element toward the right when $n_{x}$ is positive, one mesh element toward the left when $n_{x}$ is negative, and no translation occurs when $n_{x}$ is zero.
The same process is followed for $n_{y}$ with top or bottom translation.
Then, the interested DP is updated to $c \rightarrow c^{*}$ and the stencil-centric point $c$ is set to the centroid mesh element $i$ of the parallel-shifted $3 \times 3$ mesh element systems (Figure \ref{fig_stencil_shift}-b).

According to Figure \ref{fig_stencil_shift}-b, a physical quantity at $c^{*}$ is approximated by using the $\chi_{c} = 1$ MLS reconstruction corresponding to the parallel-shifted stencil-centric position $c = i$: $\phi_{c^{*}} = \phi_{c} + \mathbf{p}_{c^{*}} \cdot \mathbf{\Phi}_{c}$, similar to Eq. (\ref{eq_intrp_chi_1}).
Furthermore, the $x$-component of the first-order spatial derivative vector can be written as $\partial_{x} \phi_{c^{*}} = \partial \mathbf{p}_{c^{*}} / \partial x \cdot \mathbf{\Phi}_{c}$.
We also obtain $\partial_{y} \phi_{c^{*}}$ using the same procedure.

\section{Numerical experiments} \label{numerical_test}
\subsection{Flow past multiple stationary cylinders}
A numerical example was performed for a 2D flow past stationary multiple cylinders.
The main advantage of the proposed MCD method is that it reduces the computational cost of existing particle methods.
The mesh constraint idea in the MCD method maintains the uniformity of the DP distribution and allows computational stencils to be compact.
To investigate this feature, two types of stencils were introduced for the MLS reconstruction: the original $3 \times 3$ stencil and the expanded $9 \times 9$ stencil.
In the latter case, the influence radius of the MLS reconstruction was set to $r_e=3.5l_0$, assuming a least-squares MPS scheme with the type-A formulation \cite{tamai_LeastSquaresMoving_2014}.

\begin{figure}[H]
  \begin{center}
    \includegraphics[width=.65\linewidth]{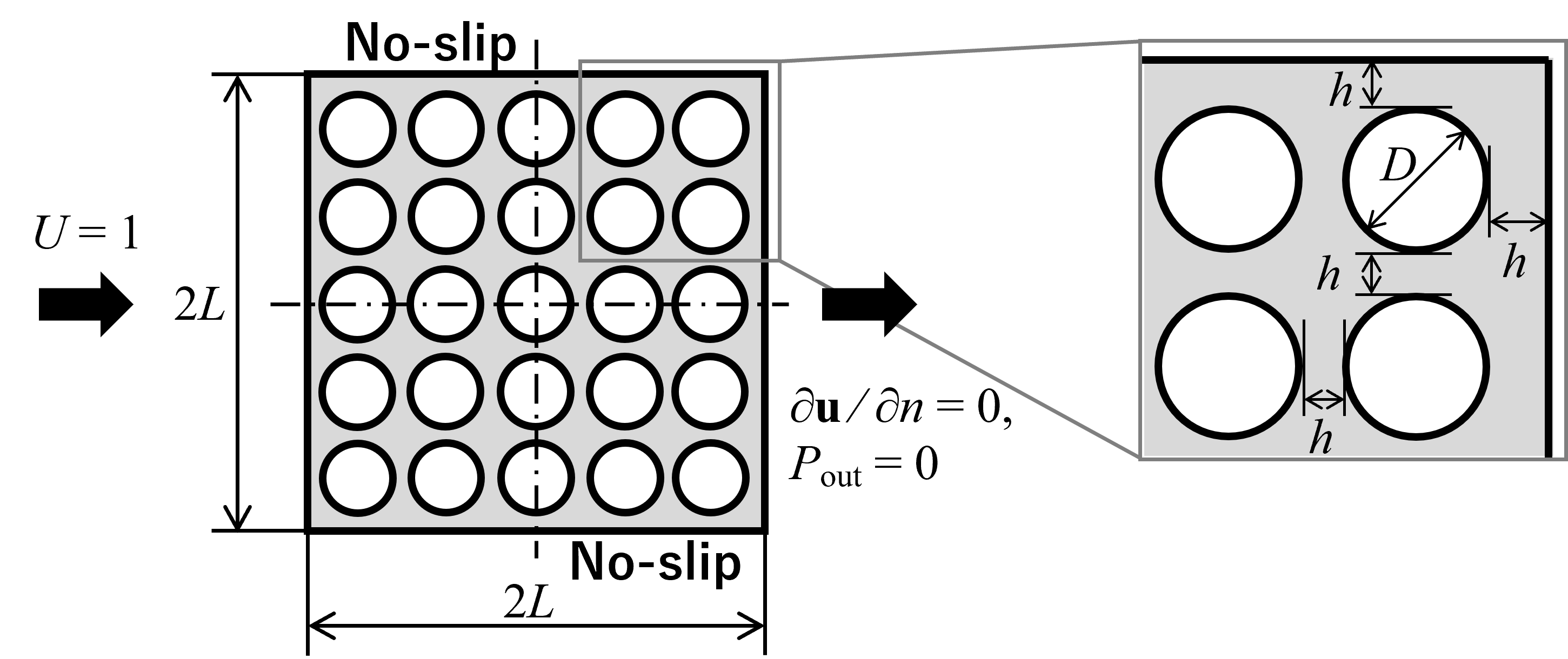}
    \caption{%
      Analysis configurations for the 2D flow past multiple stationary cylinders.
    }
    \label{fig_anal_config_multi_stat_cyl}
  \end{center}
\end{figure}

The multiple cylinders were regularly located in a square domain (Fig. \ref{fig_anal_config_multi_stat_cyl}).
The inlet velocity $U=1$ was imposed at the left boundary and the outflow condition was imposed at the right boundary of the domain.
The no-slip condition was imposed on the cylinder surfaces and the upper and lower walls.
Here, the domain size was set to $L=1$, the width of the background mesh was set to $l_0=1/64$ and the time interval was set to $\Delta t=6.1\times10^{-7}$.
The number of cylinders was set to $N_{\rm cyl}=25$. Here, approximately creeping flow was considered ($Re=0.1$).
The analyses were performed by changing the cylinder diameter $D_{\rm cyl}$ so that the inter-cylinder gap $h$ was between $h=12l_0$--$2l_0$, and steady flow fields were obtained.
The convergence criterion for the linear system of the pressure Poisson equation was set to $10^{-2}$.

\begin{figure}[H]
  \begin{center}
    \includegraphics[width=\linewidth]{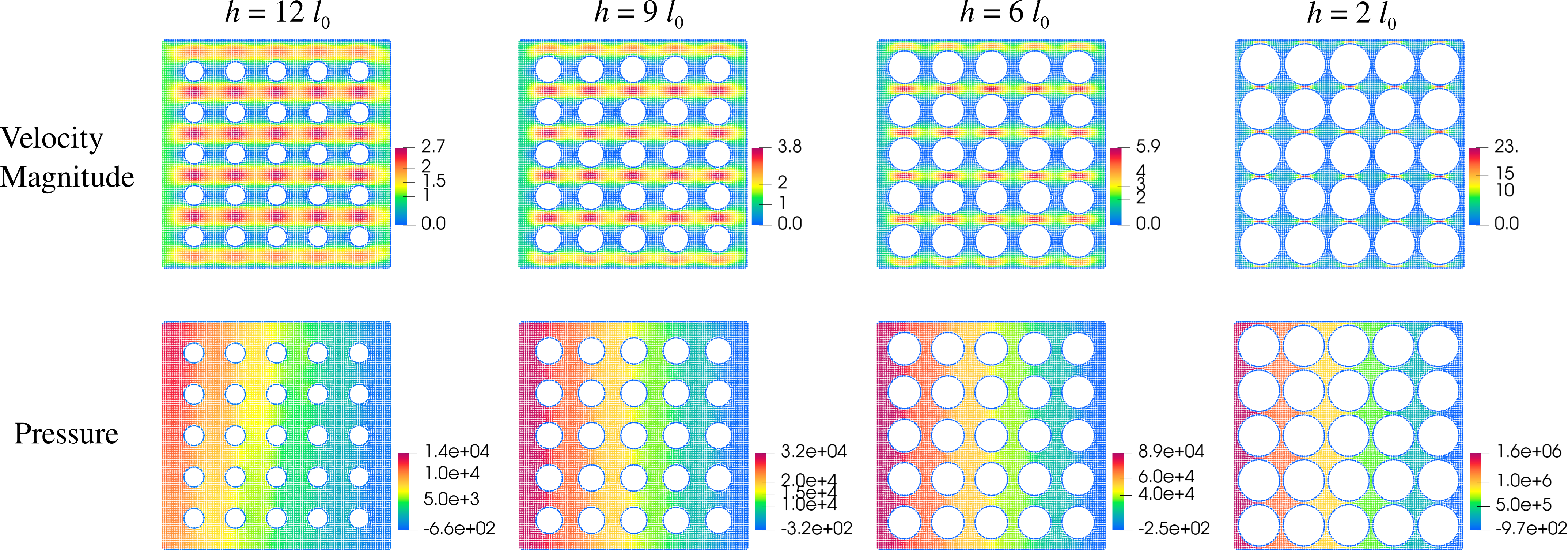}
    \caption{%
      Velocity and pressure fields calculated by MCD method using the original $3 \times 3$ stencil.
    }
    \label{fig_darcy_flowfields}
  \end{center}
\end{figure}

Fig. \ref{fig_darcy_flowfields} shows the velocity and pressure fields using the original $3 \times 3$ stencil at $h=12l_0$, $9l_0$, $6l_0$, and $2l_0$, respectively.
Smooth profiles of the solution were obtained even at the smallest gap $h=2l_0$.

\begin{figure}[H]
  \begin{center}
    \includegraphics[width=.65\linewidth]{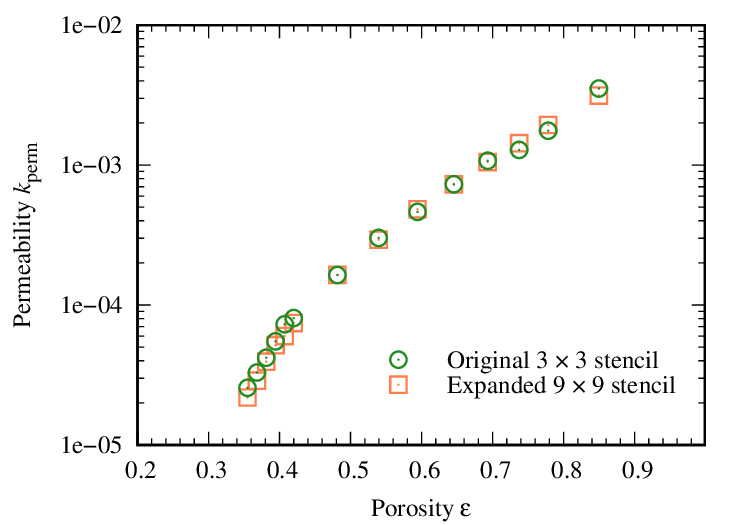}
    \caption{%
        Comparison of the relationship between permeability $k_{\rm perm}$ and porosity $\varepsilon$.
        Green and orange circles indicate the numerical results of the MCD method using the original $3 \times 3$ stencil and expanded $9 \times 9$ stencil, respectively.
    }
    \label{fig_darcy_permeability}
  \end{center}
\end{figure}

Fig. \ref{fig_darcy_permeability} shows the relationship of the porosity $\varepsilon$ and permeability $k_{\rm perm}$ defined as
\begin{equation}
  \varepsilon = 1 - \frac{\pi (D_{\rm cyl} / 2)^{2}N_{\rm cyl}}{(2L)^{2}} = 1 - \frac{\pi D^{2}_{\rm cyl}N_{\rm cyl}}{16L^{2}},
  \label{eq_darcy_porosity}
\end{equation}
\begin{equation}
  k_{\rm perm} = \frac{2 \mu U L}{\Delta P},
  \label{eq_darcy_permeability}
\end{equation}
where $\Delta P = p_{\rm in} - p_{\rm out}$ is the pressure difference between the inlet and outlet of the domain.
The results for the original $3 \times 3$ stencil and expanded $9 \times 9$ stencil were nearly identical.

Table \ref{table_darcy} shows comparisons of the number of iterations required for convergence of the pressure Poisson equation (PPE) and normalized execution time until the flow reaches steady state between the cases using the original $3 \times 3$ stencil and expanded $9 \times 9$ stencil for the inter-cylindrical gap $h=2l_0$ and $12l_0$.
The execution time is the time elapsed from the start of the execution to a particular time step, which was set to the same number in all cases.
The number of PPE iterations is given as the approximate average value for the execution.
The solution with the expanded $9 \times 9$ stencil reduces the number of PPE iterations by approximately half compared to the solution with the original $3 \times 3$ stencil.
This is believed to be the result of the wide stencil discretization communicating more information in one iteration that reduces the overall number of iterations required to converge.
Meanwhile, the total execution time was inversely related, with the original $3 \times 3$ stencil being roughly half that of the expanded $9 \times 9$ stencil.
This agrees with the fact that the number of algebraic calculations becomes small in compact stencil schemes.
In general, the particle and meshless methods require wide stencils to reconstruct a polynomial that is likely to increase computational cost compared to the grid-based methods.
The proposed MCD method can alleviate this problem by applying the mesh constraint idea.
\begin{table}[H]
  \begin{center}
    \caption{
        Comparisons of the number of iterations required for convergence of the pressure Poisson equation (PPE) and normalized execution time until the flow reaches steady state between the cases using the original $3 \times 3$ stencil and expanded $9 \times 9$ stencil for the inter-cylindrical gap $h=2l_0$ and $12l_0$. $\hat{t}$ is normalized to the execution time for the original $3 \times 3$ stencils at $h=12l_0$.
    }
    \begin{tabular}{llll} \hline
      ~~          & ~~                                & Original $3 \times 3$ stencil ~~ & Expanded $9 \times 9$ stencil  \\ \hline
      $h=12l_{0}$ & Iterations of PPE (approx.) ~~    & 120                        & 60                           \\
      ~~          & Total execution time $\hat{t}$ ~~ & 1                          & 2.63                         \\ \hline
      $h=2l_{0}$  & Iterations of PPE (approx.) ~~    & 150                        & 80                           \\
      ~~          & Total execution time $\hat{t}$ ~~ & 0.697                      & 1.16                         \\ \hline
    \end{tabular}
    \label{table_darcy}
  \end{center}
\end{table}

\subsection{Flow past a stationary cylinder}
\begin{figure}[H] 
  \begin{center}
    \includegraphics[width=.5\linewidth]{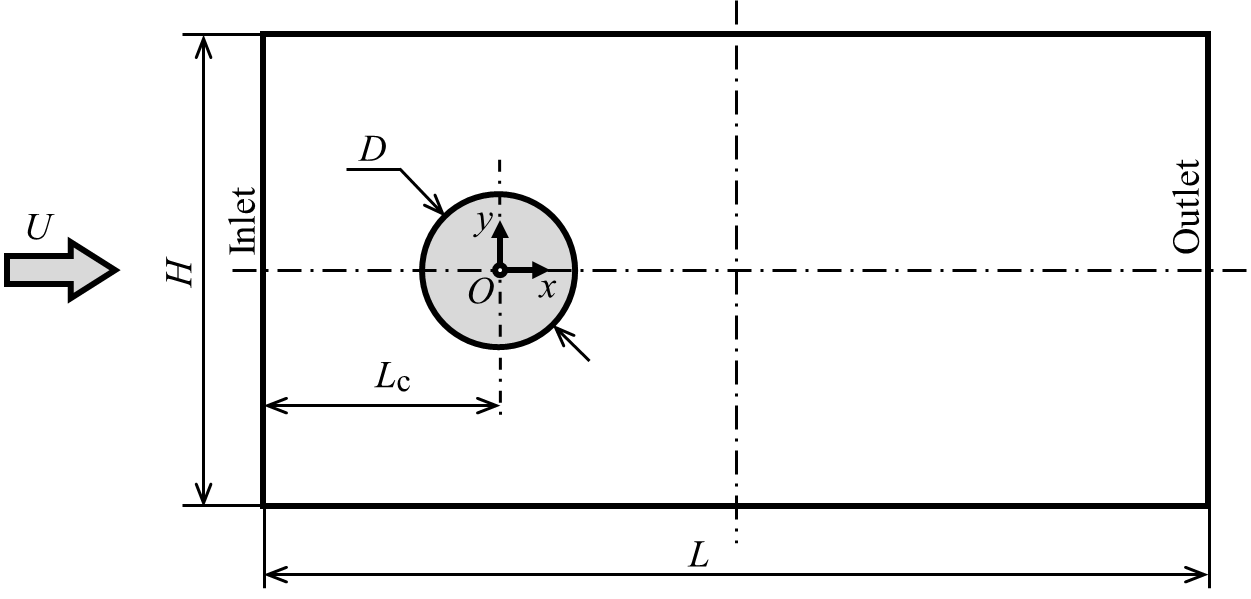}
    \caption{%
      Schematic configuration of the analysis domain for flow past a stationary cylinder.
    }
    \label{fig_anal_config_stat_cyl}
  \end{center}
\end{figure}

\begin{figure}[H] 
  \begin{center}
    \includegraphics[width=.8\linewidth]{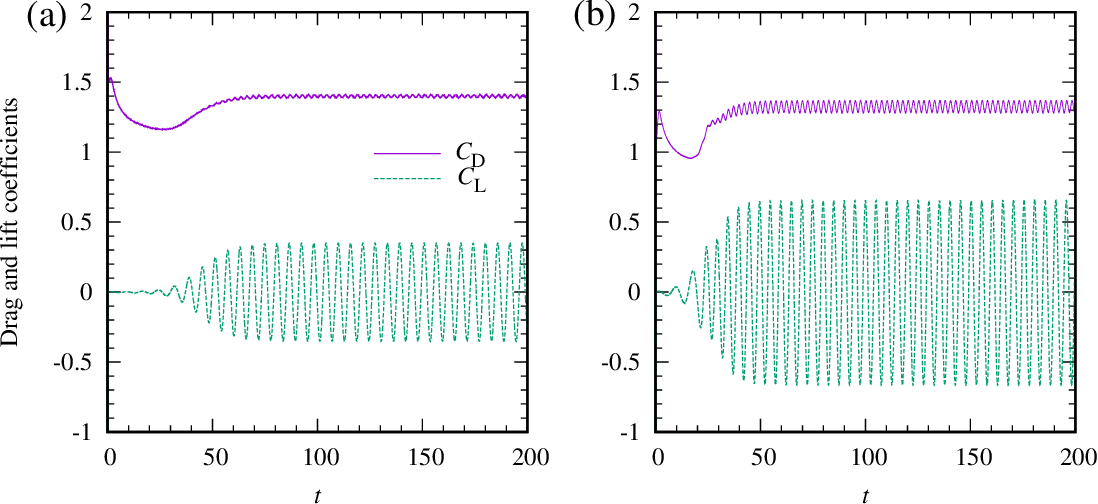}
    \caption{%
        Time histories of the drag and lift coefficients obtained by the MCD method in a problem for flow past a cylinder for $Re = 100$ (a) and $Re = 200$ (b).
        purple indicates $C_{\rm D}$ and dashed green indicates $C_{\rm L}$.
    }
    \label{fig_cd_cl_karman_Re100_Re200}
  \end{center}
\end{figure}

To check the validity of the proposed method for finite inertia flows, a numerical run was conducted for flow past a stationary cylinder.
The analysis domain is shown in Figure \ref{fig_anal_config_stat_cyl}.
The width and length of the domain were set to $H=30D$ and $L=70D$, and the distance between the center of the cylinder and the left boundary was set to $L_{\rm{c}}=20D$.
The left boundary was set as an inlet ($u=U,~v=0$), the right boundary as an outlet ($\partial \mathbf{u} / \partial n = 0,~ P = 0$), and the top and bottom boundaries and cylinder surface as a fixed wall imposing a no-slip boundary condition ($u=0,~v=0$).
The constant inlet velocity was set to $U=1$, and the diameter of the cylinder was set to $D=1$.
The kinematic viscosity $\nu$ was determined using the Reynolds number,
\begin{equation}
  Re = \frac{UD}{\nu},
  \label{eq_Re}
\end{equation}
under two conditions: $Re=100$ and $Re=200$.
The width of the background mesh (or reference spatial resolution) was set to $l_{0} = 0.051$, and the time interval was set to $\Delta t = 0.01$.
The Bi-CGSTAB method was applied to solve the linear system of the pressure Poisson equation, where the convergence criteria for the relative value of the residual norm to the initial norm were set to $1.5 \times 10^{-4}$ for both cases of $Re = 100$ and $Re = 200$.
The drag and lift coefficients $C_{\rm{D}}$ and $C_{\rm{L}}$ were estimated using the fluid force $\mathbf{F} = (F_{x},~ F_{y})$ given in Eqs. (\ref{eq_discretized_fluid_force}) and (\ref{eq_surface_area_element}) as follows:
\begin{equation}
  C_{\rm{D}} = \frac{2F_{x}}{D U^{2}},
  \label{eq_cd}
\end{equation}
\begin{equation}
  C_{\rm{L}} = \frac{2F_{y}}{D U^{2}}.
  \label{eq_cl}
\end{equation}
Figure \ref{fig_cd_cl_karman_Re100_Re200} shows the time changes of $C_{\rm D}$ and $C_{\rm L}$ for each Reynolds number.
Good periodicity was achieved without remarkable numerical oscillation.
Table \ref{table_cd_cl_st} shows the comparisons of $C_{\rm{D}}$, $C_{\rm{L}}$, and the Strouhal number $St=fD/U$ with the vortex shedding frequency $f$.
The fast Fourier transform was applied to the time histories of $C_{\rm D}$ and $C_{\rm L}$ in $t\in[100,200]$ for the calculation of the mean and standard deviation, and for frequency $f$ (using the peak frequency).
Compared with the result obtained by Liu et al. \cite{liu_PreconditionedMultigridMethods_1998} using a finite difference method with boundary-fitted mesh systems, the mean value of $C_{\rm D}$ is slightly overpredicted by approximately 4\% for $Re=100$ and 1\% for $Re=200$, and $C_{\rm L}$ was slightly different by approximately 1\% for $Re=100$ and $-$8\% for $Re=200$.
Despite this, the results were in good agreement with previous results of the immersed interface method \cite{xu_ImmersedInterfaceMethod_2006}, the method based on the immersed boundary method (IBM) \cite{cai_MovingImmersedBoundary_2017,ghomizad_SharpInterfaceDirectforcing_2021}, and least-squares-based meshfree finite difference method \cite{ding_NumericalSimulationFlows_2007}.
Moreover, our results reflect the effect of the Reynolds number more appropriately than the results in \cite{marrone_AccurateSPHModeling_2013}, which are calculated by using physically consistent $\delta$-SPH method.
\begin{table}[H]
  \begin{center}
    \caption{
      Comparison of the drag and lift coefficients of the proposed method with those of existing methods for flow past a cylinder.
      Results with * were taken from experimental studies.
    }
    \begin{tabular}{cllll} \hline
  ~~         & ~~                                                                              & $C_{\rm{D}}$      & $C_{\rm{L}}$ & $St$  \\ \hline
  $Re = 100$ & Prandtl et al.$^*$ \cite{prandtl_ErgebnisseAerodynamischenVersuchsanstalt_2009} &  1.43             & --           & --    \\
  ~~         & Williamson$^*$ \cite{williamson_ObliqueParallelModes_1989}                      & --                & --           & 0.164 \\
  ~~         & Liu et al. \cite{liu_PreconditionedMultigridMethods_1998}                       & $1.350 \pm 0.012$ & $\pm 0.339$  & 0.165 \\
  ~~         & Xu and Wang \cite{xu_ImmersedInterfaceMethod_2006}                              & $1.423 \pm 0.013$ & $\pm 0.34$   & 0.171 \\
  ~~         & Cai et al. \cite{cai_MovingImmersedBoundary_2017}                               & $1.380 \pm 0.010$ & $\pm 0.343$  & 0.160 \\
  ~~         & Ghomizad et al. \cite{ghomizad_SharpInterfaceDirectforcing_2021}                & $1.401 \pm 0.011$ & $\pm 0.344$  & 0.162 \\
  ~~         & Ding et al. \cite{ding_NumericalSimulationFlows_2007}                           & $1.356 \pm 0.010$ & $\pm 0.287$  & 0.166 \\
  ~~         & Marrone et al.  \cite{marrone_AccurateSPHModeling_2013}                         & $1.36 \pm 0.01$   & $\pm 0.24$   & 0.168 \\
  ~~         & Proposed method                                                                 & $1.399 \pm 0.009$ & $\pm 0.342$  & 0.170 \\ \hline
  $Re = 200$ & Prandtl et al.$^*$ \cite{prandtl_ErgebnisseAerodynamischenVersuchsanstalt_2009} &  1.29             & --           & --    \\
  ~~         & Williamson$^*$ \cite{williamson_ObliqueParallelModes_1989}                      & --                & --           & 0.197 \\
  ~~         & Liu et al. \cite{liu_PreconditionedMultigridMethods_1998}                       & $1.31 \pm 0.049$  & $\pm 0.69$   & 0.192 \\
  ~~         & Xu and Wang \cite{xu_ImmersedInterfaceMethod_2006}                              & $1.42 \pm 0.04$   & $\pm 0.66$   & 0.202 \\
  ~~         & Cai et al. \cite{cai_MovingImmersedBoundary_2017}                               & $1.355 \pm 0.042$ & $\pm 0.677$  & 0.200 \\
  ~~         & Ghomizad et al. \cite{ghomizad_SharpInterfaceDirectforcing_2021}                & $1.365 \pm 0.044$ & $\pm 0.687$  & 0.201 \\
  ~~         & Ding et al. \cite{ding_NumericalSimulationFlows_2007}                           & $1.348 \pm 0.050$ & $\pm 0.659$  & 0.196 \\
  ~~         & Marrone et al.  \cite{marrone_AccurateSPHModeling_2013}                         & $1.38 \pm 0.05$   & $\pm 0.68$   & 0.200 \\
  ~~         & Proposed method                                                                 & $1.325 \pm 0.036$ & $\pm 0.634$  & 0.200 \\ \hline
    \end{tabular}
    \label{table_cd_cl_st}
  \end{center}
\end{table}

\subsection{Single cylinder oscillation} \label{single_cyl_oscill}
\begin{figure}[H] 
  \begin{center}
    \includegraphics[width=.325\linewidth]{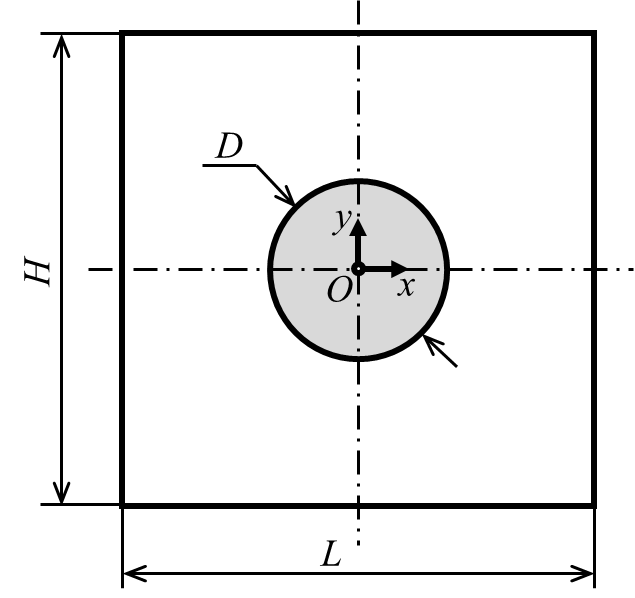}
    \caption{%
      Initial condition for the single cylinder oscillation problem, where $H=L=8D$.
    }
    \label{fig_anal_config_single_cyl_oscil}
  \end{center}
\end{figure}

A moving boundary problem for an oscillating single circular cylinder \cite{dutsch_LowReynoldsnumberFlowOscillating_1998,guilmineau_NUMERICALSIMULATIONVORTEX_2002,liao_SimulatingFlowsMoving_2010,cai_MovingImmersedBoundary_2017,chi_DirectionalGhostcellImmersed_2020,ghomizad_SharpInterfaceDirectforcing_2021} was solved.
The initial condition is shown in Figure \ref{fig_anal_config_single_cyl_oscil}.
The $x$-component of cylinder displacement $x_{\rm{c}}$ is given as 
\begin{equation}
  x_{\rm{c}} = - A \sin{(2 \pi f t)},
\end{equation}
where $A$ is the amplitude of cylinder oscillation and $f$ is the frequency.
The Reynolds number is defined as Eq. (\ref{eq_Re}), and the Keulegan-Carpenter number $KC$ is defined as
\begin{equation}
  KC = \frac{U}{f D},
\end{equation}
where $D$ is the cylinder diameter and the reference velocity $U$ is assumed to be the maximum speed of the cylinder.
To calculate flow for $Re = 100$ and $KC = 5$, parameters were set to $D=1$, $f = 1$, $A = 5 / (2 \pi f)$, and $U = 2 \pi f A$.
The width of the background mesh was set to $l_{0} = 2.55 \times 10^{-2}$ and the time interval to $\Delta t = 1/900$.
The no-slip condition was imposed on the cylinder and the outflow condition ($\partial \mathbf{u} / \partial n = 0$, $p = 0$) on the left, right upper, and lower walls.
The convergence criterion value of the Bi-CGSTAB method for solving the linear system of the pressure Poisson equation was set to $10^{-5}$.

\begin{figure}[H] 
  \begin{center}
    \includegraphics[width=.85\linewidth]{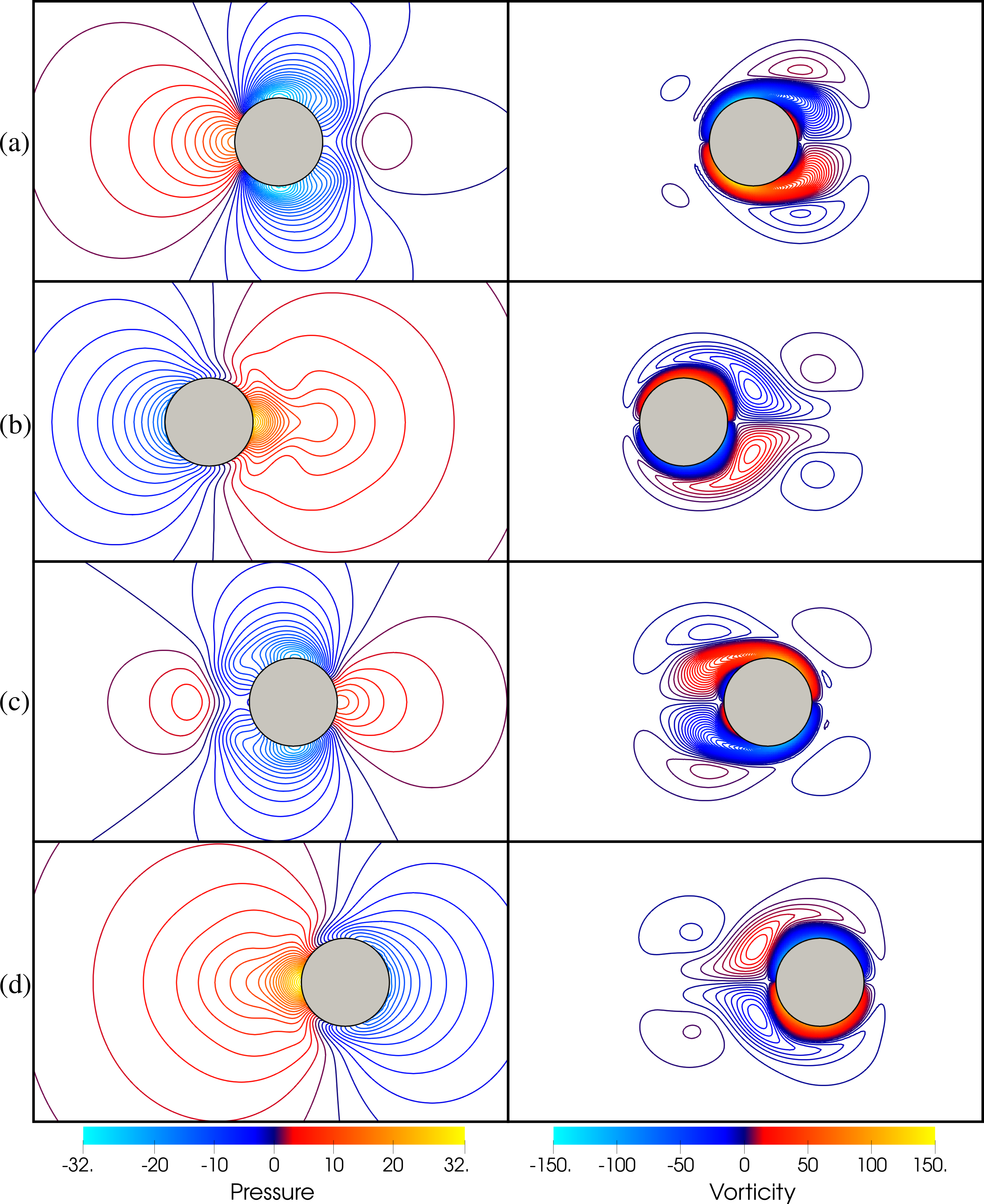}
    \caption{%
        Numerical results for the pressure (left) and vorticity (right) isolines around an oscillating cylinder.
        The phase angles of the cylinder position were (a) 0$^{\circ}$, (b) 96$^{\circ}$, (c) 192$^{\circ}$, and (d) 288$^{\circ}$.
        The contour levels of the pressure were $-32$ to 32 by increments of 1.28, and the contour levels of the vorticity were $-150$ to 150 by increments of 2.4.
    }
    \label{fig_contour_pressure_vorticity_single_cyl_oscil}
  \end{center}
\end{figure}

\begin{figure}[H] 
  \begin{center}
    \includegraphics[width=\linewidth]{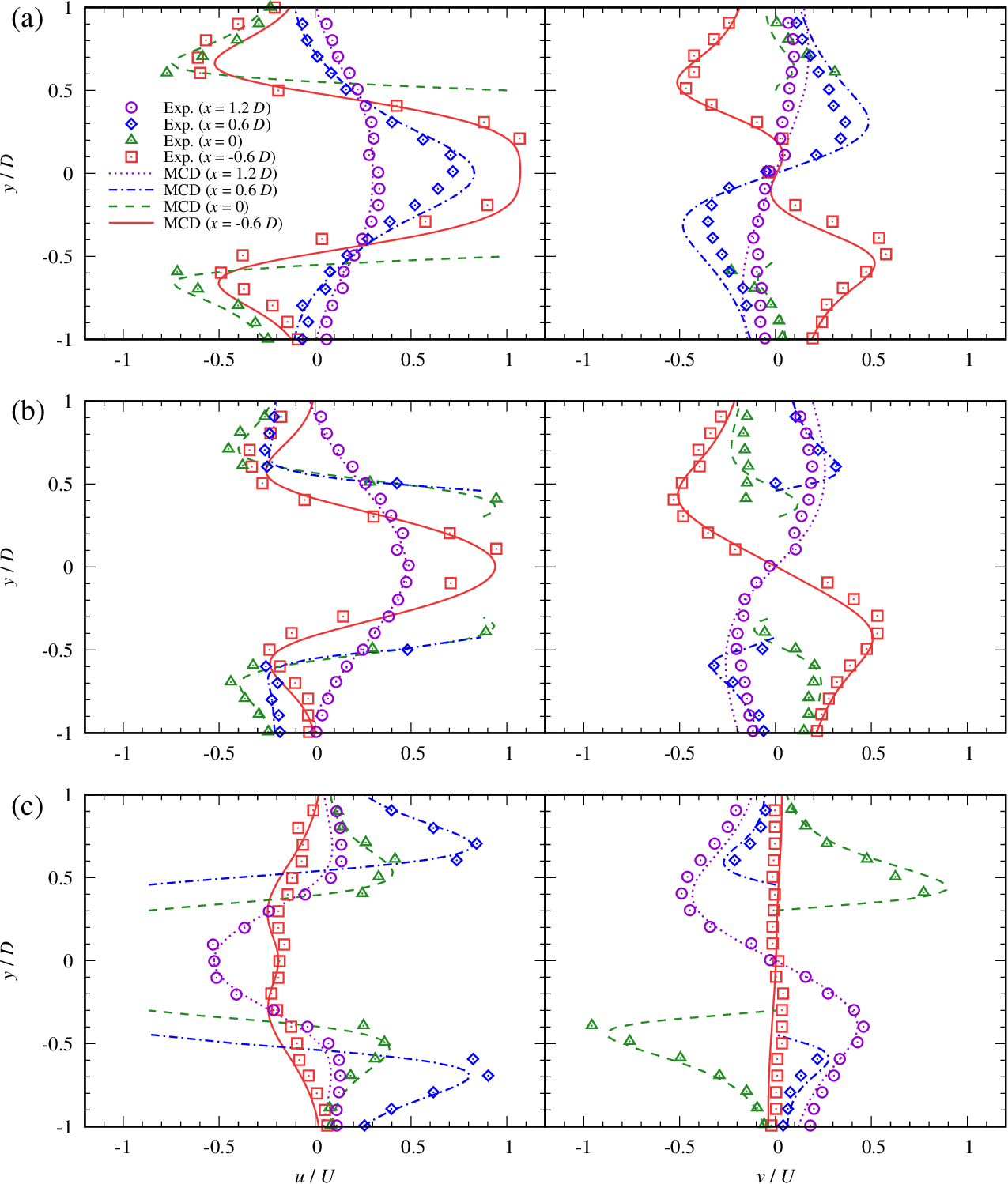}
    \caption{%
        Comparisons of $x$ and $y$ components of velocities at four cross-sections: $x=-0.6D$ (red line/square), $x=0$ (green dashed line/triangle), $x=0.6D$ (blue dash-dotted line/diamond), and $x=1.2D$ (purple dotted line/circle).
        The phase angles of the cylinder position were (a) 180$^\circ$, (b) 210$^\circ$, and (c) 330$^\circ$.
        Experimental results were taken from \cite{dutsch_LowReynoldsnumberFlowOscillating_1998}.
    }
    \label{fig_vel_cross_section_single_cyl_oscil}
  \end{center}
\end{figure}

\begin{figure}[H] 
  \begin{center}
    \includegraphics[width=.6\linewidth]{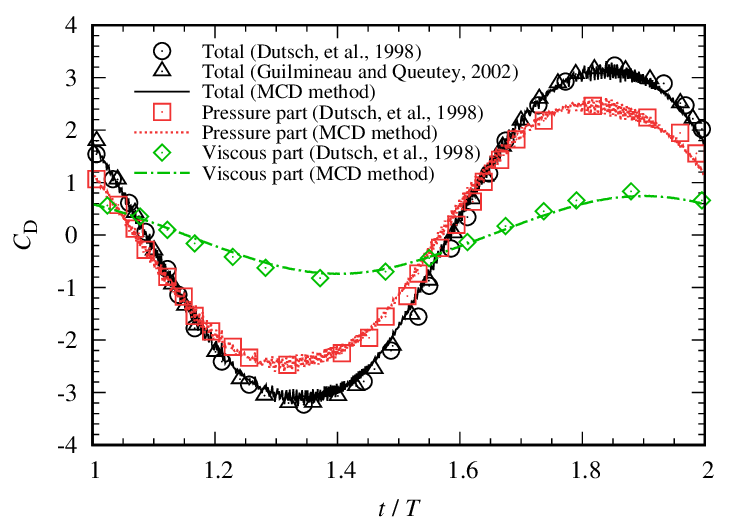}
    \caption{%
        Comparison of the time variation of the drag force coefficient and its components.
        $T$ is the period of cylinder motion.
        The circle/triangle/square/diamond symbols indicate previous results \cite{dutsch_LowReynoldsnumberFlowOscillating_1998,guilmineau_NUMERICALSIMULATIONVORTEX_2002}.
    }
    \label{fig_cd_single_cyl_oscil_total_pres_visc}
  \end{center}
\end{figure}

Figure \ref{fig_contour_pressure_vorticity_single_cyl_oscil} shows instantaneous contours of pressure and vorticity at four different times for the following phase angles: (a) 0$^{\circ}$, (b) 96$^{\circ}$, (c) 192$^{\circ}$, and (d) 288$^{\circ}$.
A symmetrical vortex pair was shedding at the rear side of the direction of motion, and an asymmetric pressure field before and after the moving cylinder appeared, as shown by D\"utsch et al. \cite{dutsch_LowReynoldsnumberFlowOscillating_1998}.
Figure \ref{fig_vel_cross_section_single_cyl_oscil} shows the $x$ and $y$-components of the velocity on the four cross-sections at $x = -0.6D$, 0, $0.6D$, $1.2D$, and three phase angles: (a) 180$^{\circ}$, (b) 210$^{\circ}$, and (c) 330$^{\circ}$.
Velocity fields in the vicinity of the cylinder are well captured with the experimental results by D\"utsch et al. \cite{dutsch_LowReynoldsnumberFlowOscillating_1998}, and discrepancies with the experiment are at the same level to existing numerical results \cite{ghomizad_SharpInterfaceDirectforcing_2021,dutsch_LowReynoldsnumberFlowOscillating_1998,guilmineau_NUMERICALSIMULATIONVORTEX_2002,liao_SimulatingFlowsMoving_2010,cai_MovingImmersedBoundary_2017,chi_DirectionalGhostcellImmersed_2020}.
Figure \ref{fig_cd_single_cyl_oscil_total_pres_visc} shows the time histories of $C_{\rm D}$ and its components, pressure, and viscous parts.
Although a few numerical oscillations appeared in the result, it captured both tendencies of the experiment and numerical result well, including the inertial and viscous contributions.

The proposed MCD method involves a problem for discontinuous change of DPs due to restricting the movement of DPs, where the inner DPs assigned to describe the fluid motion suddenly appear/disappear when a moving interface crosses an edge of background mesh.
This is an unavoidable feature of the MCD method that focuses on improving computational cost, and it would be a reason causing loss of numerical conservation and associated numerical dissipation and oscillation.

\begin{figure}[H] 
  \begin{center}
    \includegraphics[width=.55\linewidth]{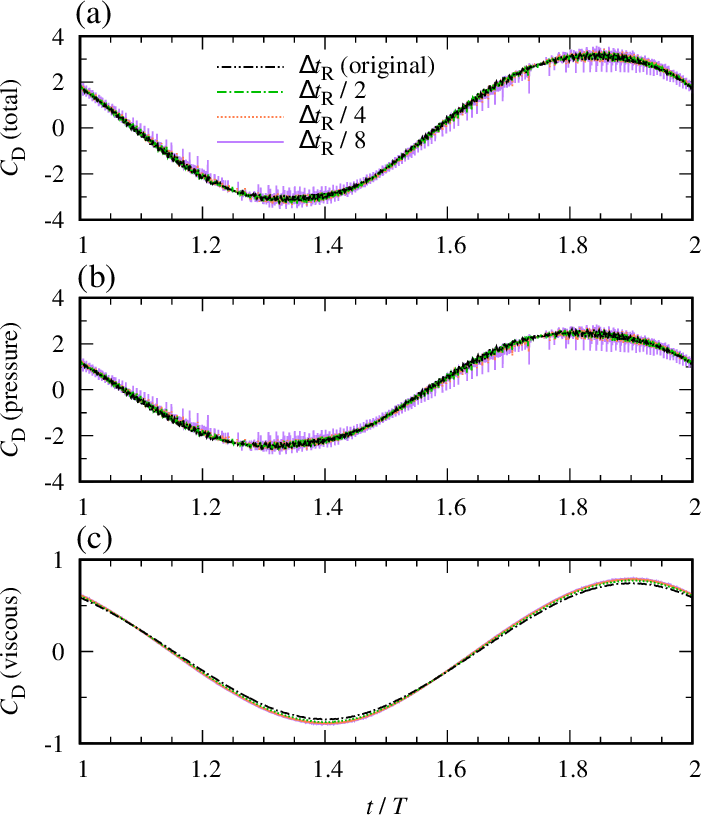}
    \caption{%
        Comparison of the time variation of the drag force coefficient and its components.
        (a) indicates $C_{\rm D}$ and (b) and (c) indicate the pressure and viscous part, respectively.
        The black dash-dot-dot line indicates the result that uses the original temporal resolution $\Delta{t}_{\rm R}$. 
        Dash-dotted green, dashed orange, and purple indicate the results for $\Delta t=\Delta{t}_{\rm R}/2$, $\Delta t=\Delta{t}_{\rm R}/4$, and $\Delta t=\Delta{t}_{\rm R}/8$, respectively.
    }
    \label{fig_cd_single_cyl_oscil_dt_decline}
  \end{center}
\end{figure}

To investigate the effect of the time resolution on fluid forces, $C_{\rm D}$ was estimated using different $\Delta t$ (Figure \ref{fig_cd_single_cyl_oscil_dt_decline}).
Non-negligible numerical oscillations appeared for the pressure part in the fluid force when $\Delta t$ decreased.
The number of updating iterations increased as $\Delta t$ decreased to reach the same physical time.
This accumulated interpolation errors in the ALE formulation and broke the conservation features of the numerical solutions.
Moreover, the proposed method applied a collocation arrangement to the pressure projection method.
This generated checkerboard oscillations for the pressure, which became remarkable when $\Delta t$ decreased.
A stable formulation to suppress the oscillation \cite{matsunaga_StabilizedLSMPSMethod_2022} is required.

\subsection{Two cylinders passing each other}
\begin{figure}[H] 
  \begin{center}
    \includegraphics[width=.55\linewidth]{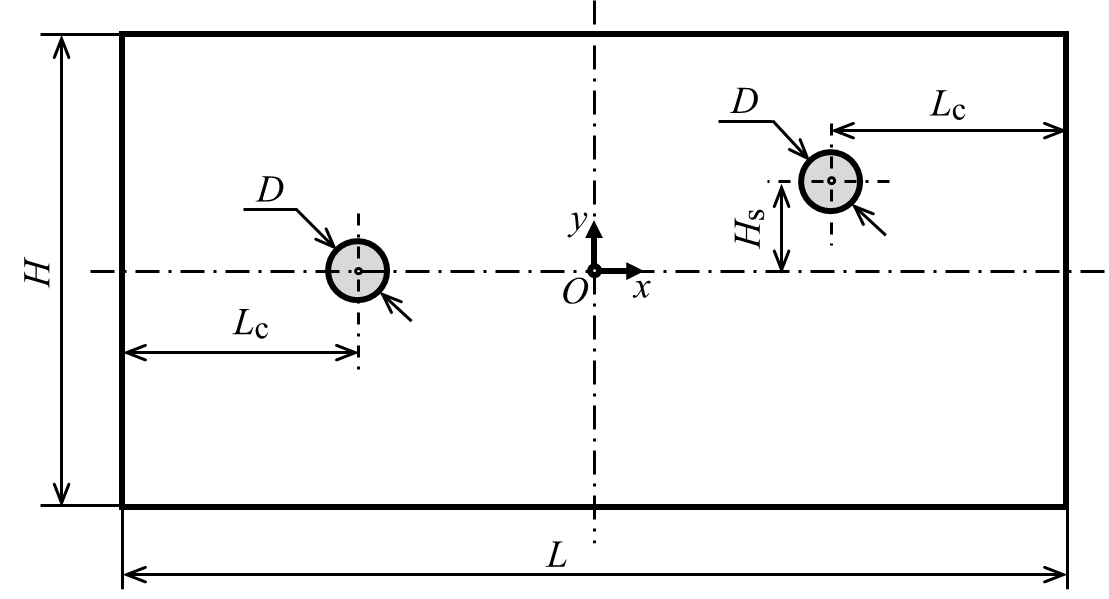}
    \caption{%
      Initial setup for the problem of two cylinders passing each other.
    }
    \label{fig_anal_config_two_cyl}
  \end{center}
\end{figure}

A numerical experiment for two cylinders passing each other was proposed in \cite{russell_CartesianGridMethod_2003} and then developed to avoid the initial fluctuation in \cite{xu_ImmersedInterfaceMethod_2006,liao_SimulatingFlowsMoving_2010}.
The initial setup is shown in Figure \ref{fig_anal_config_two_cyl}.
$D$ is the cylinder diameter, $H$ and $L$ are the width and length of the analysis domain, respectively, $L_c$ is an offset value toward the $x$-component for the lower and upper cylinder position based on the left and right side wall, and $H_s$ is an offset value toward the $y$-component for the upper cylinder.
These parameters were set as follows: $D=1$, $H=16D$, $L=32D$, $L_{\rm{c}}=8D$, and $H_{\rm{s}}=1.5D$.
In this test, to ignore the initial fluctuation, the cylinders were moved gradually as follows:
First, the lower and upper cylinders oscillated in the $x$ direction for two periods at the center of their initial position, and then each cylinder moved parallel to the $x$-axis at a constant velocity.
The centroids of each cylinder $\mathbf{x}_{\rm{lc}} = (x_{\rm{lc}},~y_{\rm{lc}})$ and $\mathbf{x}_{\rm{uc}} = (x_{\rm{uc}},~y_{\rm{uc}})$ are given by 
\begin{equation}
  x_{\rm{lc}}(t) = \left\{ \begin{array}{lr} \displaystyle
    x_{\rm{lc}}(0) + A \sin{( 2 \pi f t)},   & (0  \leq t \leq 2T), \\[8pt]
    x_{\rm{lc}}(0) + U (t - 2T),  & (2T \leq t \leq 4T),
  \end{array} \right.
  \quad y_{\rm{lc}}=0,
\end{equation}
and
\begin{equation}
  x_{\rm{uc}}(t) = \left\{ \begin{array}{lr} \displaystyle
    x_{\rm{uc}}(0) - A \sin{( 2 \pi f t)},  & (0  \leq t \leq 2T), \\[8pt]
    x_{\rm{uc}}(0) - U (t - 2T), & (2T \leq t \leq 4T),
  \end{array} \right.
  \quad y_{\rm{uc}}=H_{\rm{s}},
\end{equation}
where $A = 1 / (2 \pi f)$ is the amplitude of the cylinder oscillation, $f = 1 / T$ is the frequency with periodic time $T = 8$, and $U$ is the maximum speed of the cylinders set to $U=2\pi fA$.
The analysis domain was assumed to be closed: the left, right, top, and bottom boundaries were defined as a fixed wall ($u=0,~ v=0$).
The Reynolds number was set to $Re = UD / \nu = 40$.

The width of the background mesh was set to $l_{0} = 2.55 \times 10^{-2}$, and the time interval was set to $\Delta t = 5 \times 10^{-3}$ so that the maximum CFL number was approximately 0.2.
The no-slip condition was used for the cylinder surfaces.
The convergence criterion for solving the linear system of the pressure Poisson equation was set to $1 \times 10^{-2}$.

\begin{figure}[H] 
  \begin{center}
    \includegraphics[width=\linewidth]{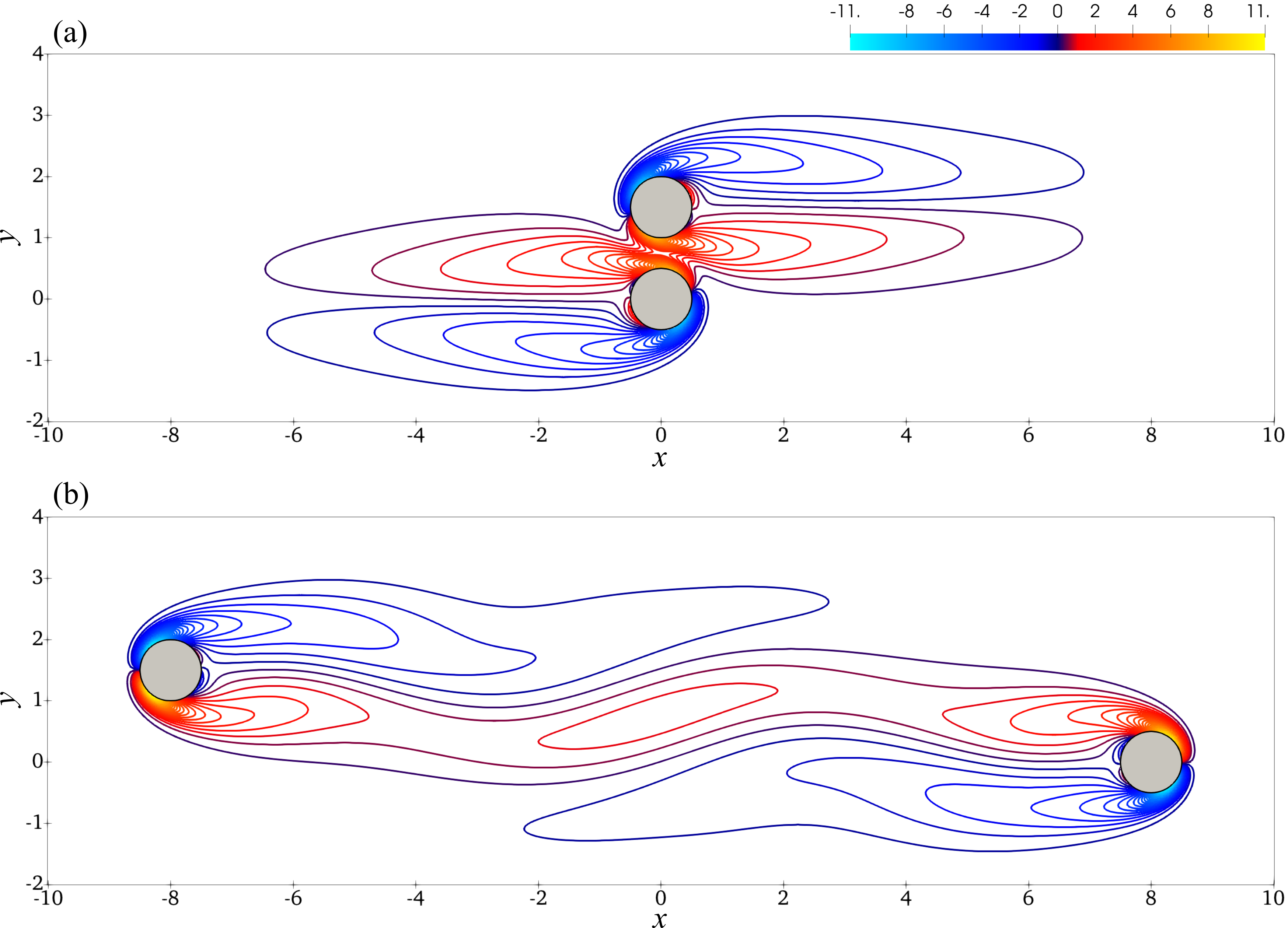}
    \caption{%
        Contours of vorticity around the two moving cylinders at $t = 3T$ (a) and $t = 4T$ (b).
        The contour levels are $-11$ to $11$ by increments of 0.4.
    }
    \label{fig_contour_vorticity_two_cyl}
  \end{center}
\end{figure}

Figure \ref{fig_contour_vorticity_two_cyl} shows isolines for vorticity fields at $t = 3T$ and $4T$.
Compared with the existing result \cite{xu_ImmersedInterfaceMethod_2006}, the structure of vortices was in good agreement with the result, and indicated that the proposed method could calculate moving boundary flow correctly with multiple bodies. 

\begin{figure}[H] 
  \begin{center}
    \includegraphics[width=.55\linewidth]{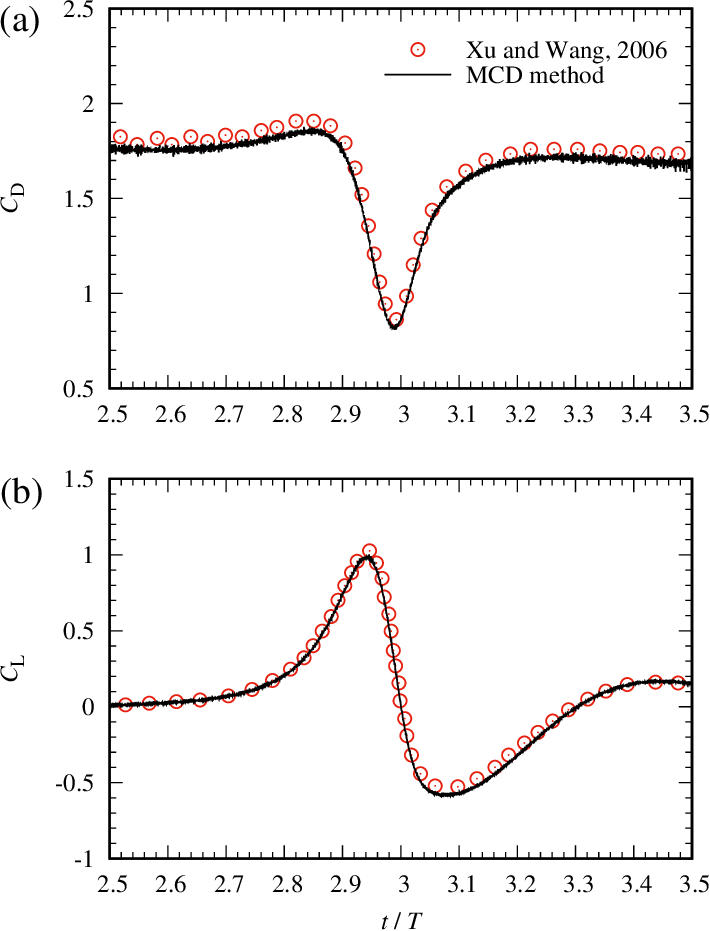}
    \caption{%
        Time changes for the drag coefficient $C_{\rm{D}}$ (a) and lift coefficient $C_{\rm{L}}$ (b).
        Time $t$ is normalized with the period of cylinder oscillation $T$.
        The black line indicates the present result and the red dots indicate the previous numerical result \cite{xu_ImmersedInterfaceMethod_2006}.
    }
    \label{fig_cd_cl_two_cyl}
  \end{center}
\end{figure}

Figure \ref{fig_cd_cl_two_cyl} shows the comparison of the time changes for the drag and lift coefficients $C_{\rm{D}}$ and $C_{\rm{L}}$, respectively, with the previous result for $t=2.5T$--$3.5T$.
Although a slight bias difference arose in $C_{\rm L}$ for the overall prediction, the time change behavior of $C_{\rm L}$ captured the previous result well \cite{xu_ImmersedInterfaceMethod_2006}.


\subsection{Influence of the spatial resolution on the gap between moving boundaries} \label{two_cyl_gap}
\begin{figure}[H] 
  \begin{center}
    \includegraphics[width=.3\linewidth]{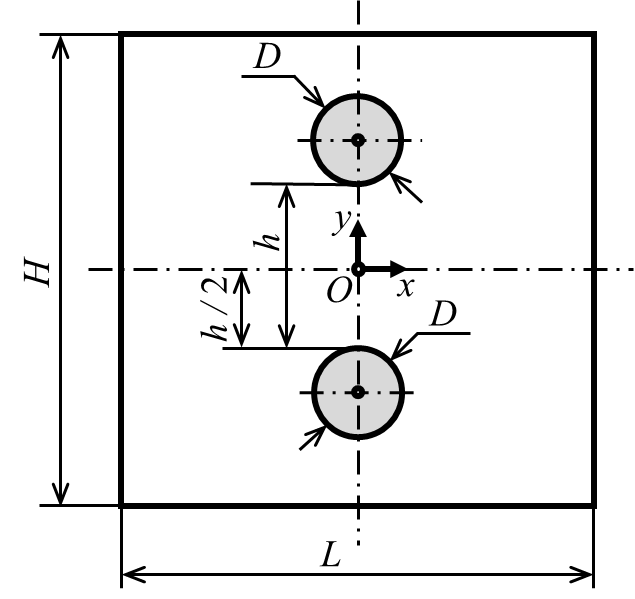}
    \caption{%
      Analysis domain for two cylinders passing each other, where the cylinders oscillate periodically in the $x$-direction in opposite phases to each other.
      The gap distance $h$ between the cylinders is varied as $h = 8 l_{0}$, $6 l_{0}$, $4 l_{0}$, $3 l_{0}$, and $2 l_{0}$.
    }
    \label{fig_anal_config_two_cyl_change_gap}
  \end{center}
\end{figure}

\begin{figure}[H] 
  \begin{center}
    \includegraphics[width=.95\linewidth]{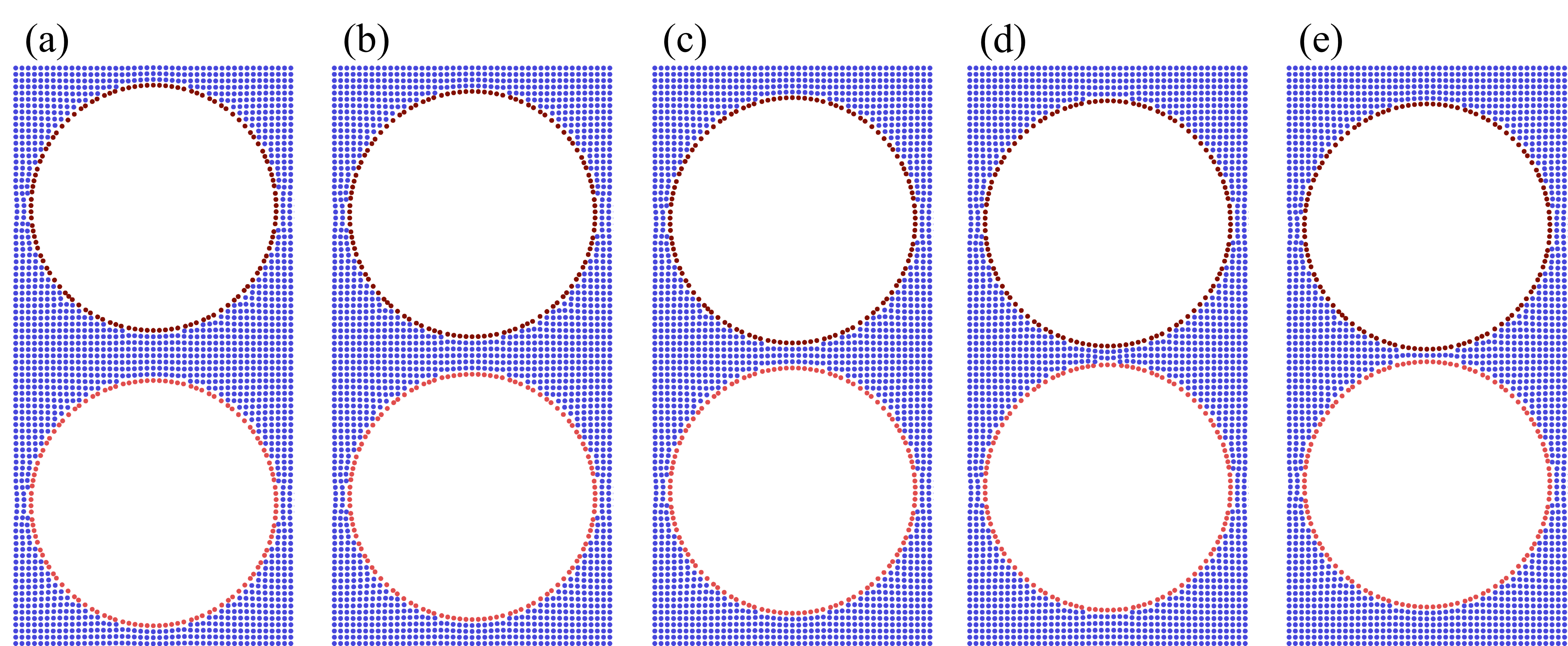}
    \caption{%
      DP arrangements near the gap when cylinders are closest to each other in the case of (a) $h = 8 l_{0}$, (b) $h = 6 l_{0}$, (c) $h = 4 l_{0}$, (d) $h = 3 l_{0}$, and (e) $h = 2 l_{0}$.
    }
    \label{fig_dp_arrangement_two_cyl_change_gap}
  \end{center}
\end{figure}

In the previous section, the calculation of fluid flows and forces acting on moving boundaries was validated by comparing our results with existing numerical solutions \cite{xu_ImmersedInterfaceMethod_2006}.
In this section, two-cylinder oscillation problems are solved with several different gaps $h$ between cylinder interfaces and the effect of the spatial resolution on numerical solutions is investigated.
An analysis domain was set as shown in Figure \ref{fig_anal_config_two_cyl_change_gap}, where $H = L = 8 D$, $D = 1$, and $h = 8 l_{0}$, $6 l_{0}$, $4 l_{0}$, $3 l_{0}$, $2 l_{0}$.
Additionally, the DP arrangements when the cylinders were closest to each other are shown in Figure \ref{fig_dp_arrangement_two_cyl_change_gap}.
The flow for $Re = 100$ was calculated.
The width of the background mesh was set to $l_{0} = 2.55 \times 10^{-2}$ and the time interval to $\Delta t = 1/900$.
All the domain boundaries were set as an outflow condition ($\partial \mathbf{u} / \partial n = 0$, $p = 0$).
The convergence criterion for the linear system of the pressure Poisson equation was set to $1 \times 10^{-5}$.
The positions of the lower and upper cylinders at time $t$ were
\begin{equation}
  x_{\rm{lc}} (t) = A \sin{(2 \pi f t)},
\end{equation}
\begin{equation}
  x_{\rm{uc}} (t) = - A \sin{(2 \pi f t)},
\end{equation}
where $A = 5 / (2 \pi f)$ and $f = 1$, similar to Section \ref{single_cyl_oscill}.

\begin{figure}[H] 
  \begin{center}
    \includegraphics[width=.68\linewidth]{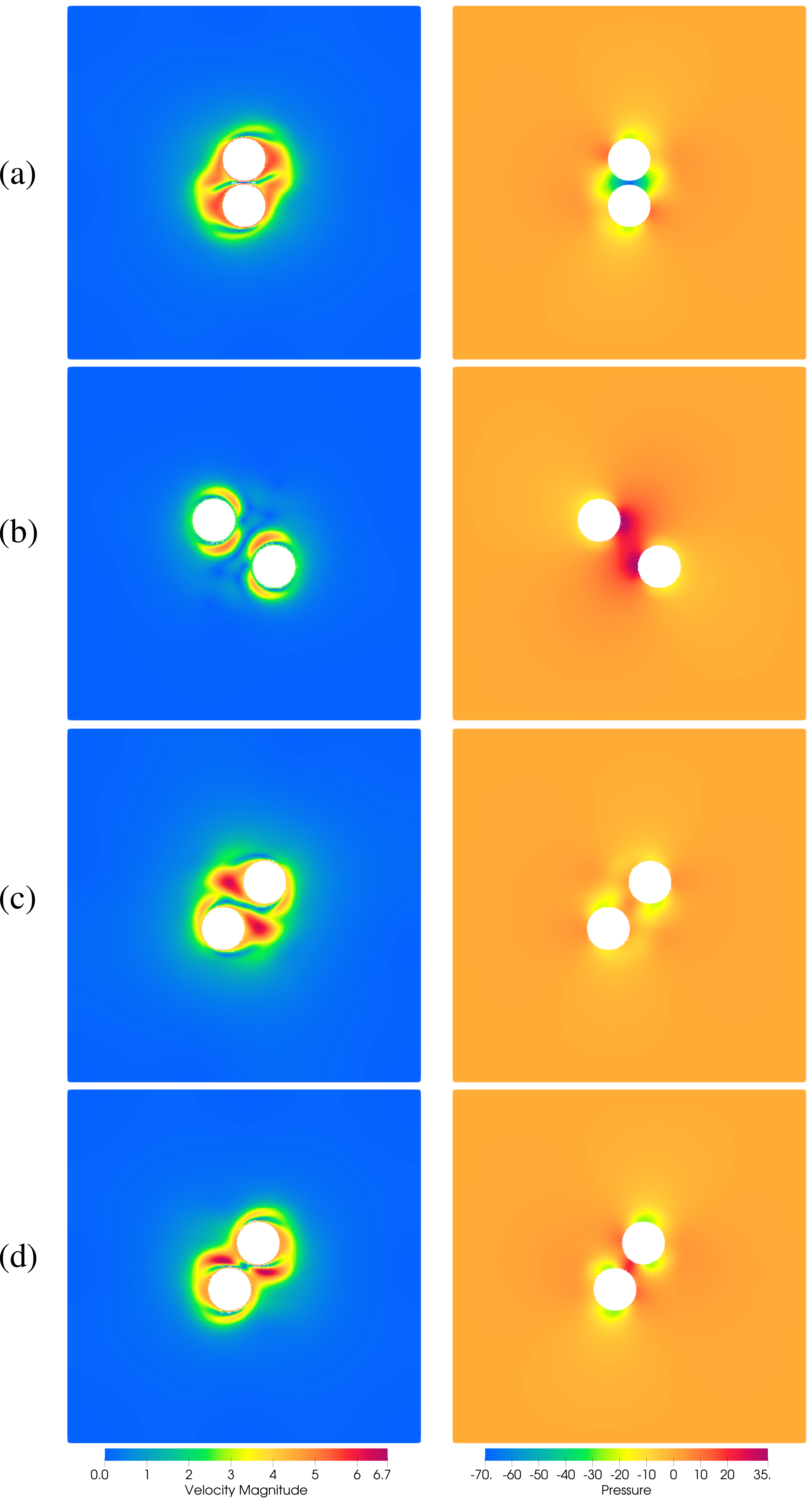}
    \caption{%
        Velocity (left) and pressure (right) fields in the case of $h = 2 l_{0}$ at the four phase angles: (a) 0$^{\circ}$, (b) 120$^{\circ}$, (c) 216$^{\circ}$, and (d) 336$^{\circ}$.
    }
    \label{fig_two_cyl_gap_velocity_pressure}
  \end{center}
\end{figure}

\begin{figure}[H] 
  \begin{center}
    \includegraphics[width=.53\linewidth]{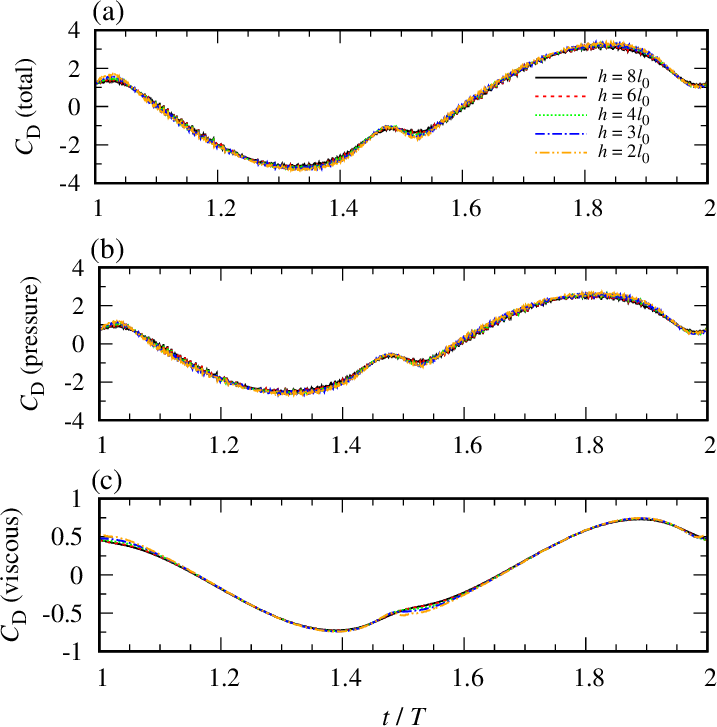}
    \caption{%
        Time variation of $C_{\rm D}$ and its components in the case of changing the gap $h$ between cylinders as $h = 8 l_{0}$, $6 l_{0}$, $4 l_{0}$, $3 l_{0}$, and $2 l_{0}$.
        (a) indicates the total $C_{\rm D}$ and (b) and (c) indicate the pressure and viscous part, respectively.
    }
    \label{fig_cd_two_cyl_gap_total_pres_visc}
  \end{center}
\end{figure}

\begin{figure}[H] 
  \begin{center}
    \includegraphics[width=.53\linewidth]{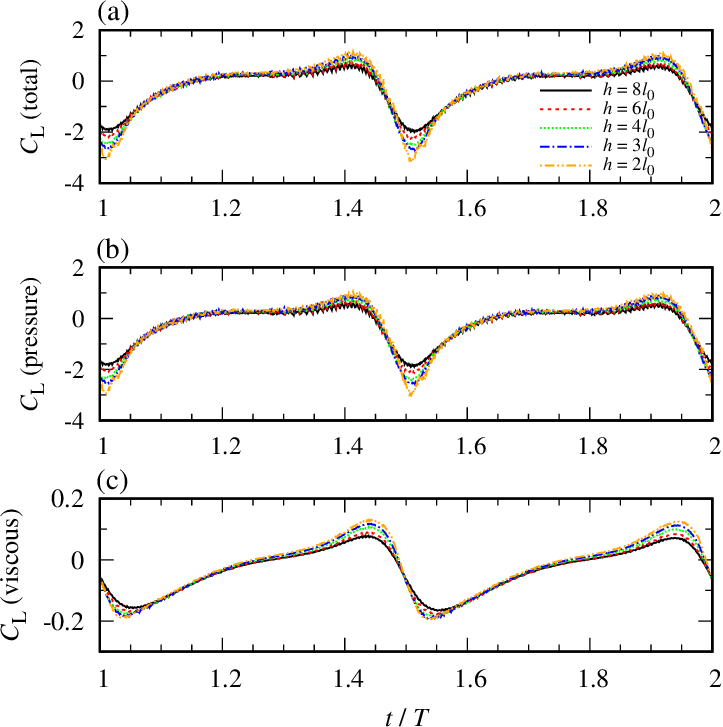}
    \caption{%
        Time variation of $C_{\rm L}$ and its components in the case of changing the gap $h$ between cylinders as $h = 8 l_{0}$, $6 l_{0}$, $4 l_{0}$, $3 l_{0}$, and $2 l_{0}$.
        (a) indicates the total $C_{\rm L}$ and (b) and (c) indicate the pressure and viscous part, respectively.
    }
    \label{fig_cl_two_cyl_gap_total_pres_visc}
  \end{center}
\end{figure}

Figure \ref{fig_two_cyl_gap_velocity_pressure} shows the velocity and pressure fields in the case of $h = 2 l_{0}$ at the four different phase angles.
The proposed method obtained both the velocity and pressure fields smoothly over each gap $h$.
Figure \ref{fig_cd_two_cyl_gap_total_pres_visc} shows the temporal change of $C_{\rm D}$ and its components (pressure and viscous parts).
Figure \ref{fig_cl_two_cyl_gap_total_pres_visc} also shows the temporal change of $C_{\rm L}$, where the result captured well the high fluctuation of the pressure-induced lift force attributed to fluid dynamics interactions between cylinders before and after they passed each other (around $t/T \approx 1.5$) as their gap $h$ decreased.
Both $C_{\rm D}$ and $C_{\rm L}$ were obtained stably at the same level as the numerical oscillation for any $h$.
Thus, the proposed method obtained fluid forces by properly reflecting the effect of wall boundaries even in situations where the boundaries are very close to each other.
These results suggest that the proposed method can stably calculate both the fluid flow and fluid force if there is at least one DP between boundaries.

\subsection{Motion of a flapping wing}
\begin{figure}[H]
  \begin{center}
    \includegraphics[width=.45\linewidth]{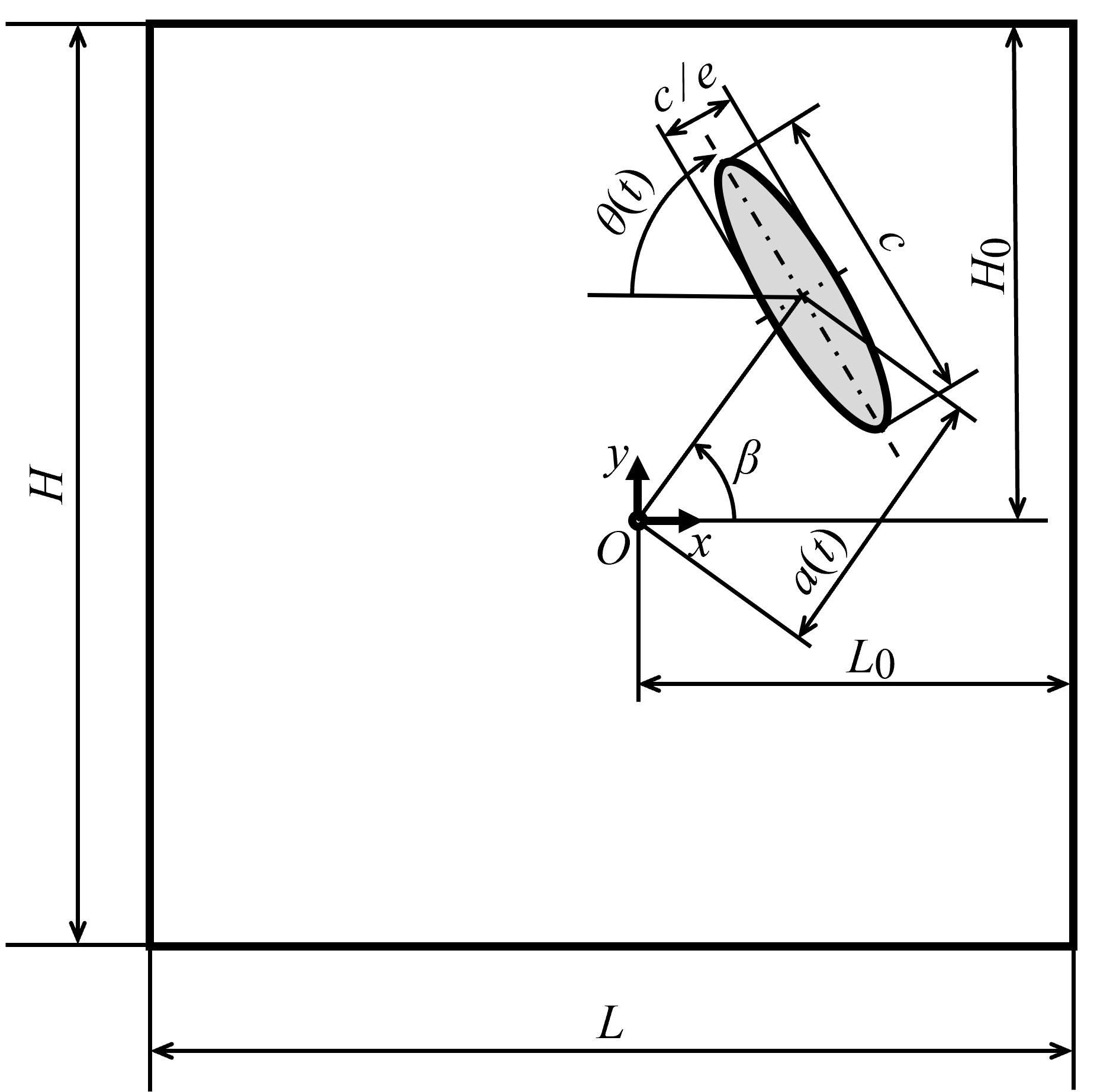}
    \caption{%
      Schematic configuration of the analysis domain and definition of parameters for the flapping wing problem \cite{xu_ImmersedInterfaceMethod_2006}, where the attack angle $\theta (t)$ and rotation center $a (t)$ of the elliptical wing are temporally changed.
    }
    \label{fig_anal_config_flapping_wing}
  \end{center}
\end{figure}

Finally, a problem was solved for complex flows with a flapping wing, initially considered in \cite{janewang_TwoDimensionalMechanism_2000} to explain the mechanism of an insect hovering and developed by several researchers \cite{janewang_TwoDimensionalMechanism_2000,xu_ImmersedInterfaceMethod_2006,sui_HybridImmersedboundaryMultiblock_2007,yang_SimpleEfficientDirect_2012,cai_MovingImmersedBoundary_2017,chi_DirectionalGhostcellImmersed_2020}.
The analysis domain is shown in Figure \ref{fig_anal_config_flapping_wing}, where $c$ denotes the chord length, $e$ denotes the aspect ratio of the elliptical wing, $H$ and $L$ denote the width and length of the analysis domain, respectively, $H_{0}$ and $L_{0}$ denote the distance from the origin to the upper and right wall boundary, respectively, and $\beta$ denotes the elevation angle of the stroke plane of the elliptical wing.
These parameters were set to $c = 1$, $e = 4$, $H = L = 16 c$, $H_{0} = L_{0} = 6 c$, and $\beta = \pi / 3$.
$\theta (t)$ denotes the attack angle and $a (t)$ denotes the offset distance from the origin to the rotation center of the elliptical wing: 
\begin{equation}
  \theta (t) = \theta_{0} \left[ 1 - \sin{\left( \frac{2t}{a_{0}} + \phi \right)} \right],
  \label{eq_theta_flapping_wing}
\end{equation}
\begin{equation}
  a (t) = \frac{a_{0}}{2} \left[ 1 + \cos{\left( \frac{2t}{a_{0}} \right)} \right],
  \label{eq_a_flapping_wing}
\end{equation}
where $\theta_{0}$ is the initial attack angle, $\phi$ is the phase angle, and $a_{0}$ is the amplitude of translational motion of the wing.
These parameters were set to $\theta_{0} = \pi / 4$, $\phi = 0$, and $a_{0} = 5 / 2$.
Note that, according to Eqs. (\ref{eq_theta_flapping_wing}) and (\ref{eq_a_flapping_wing}), the period $T$ of the elliptical wing motion is $T = \pi a_{0}$.
The width of the background mesh was set to $l_{0} = 51 / 1600$ and the time interval $\Delta t = \pi / 600$.
The no-slip condition was imposed on the elliptical cylinder and the left, right upper, and lower walls.
The convergence criterion for the linear system of the pressure Poisson equation was set to $2 \times 10^{-3}$.

\begin{figure}[H] 
  \begin{center}
    \includegraphics[width=\linewidth]{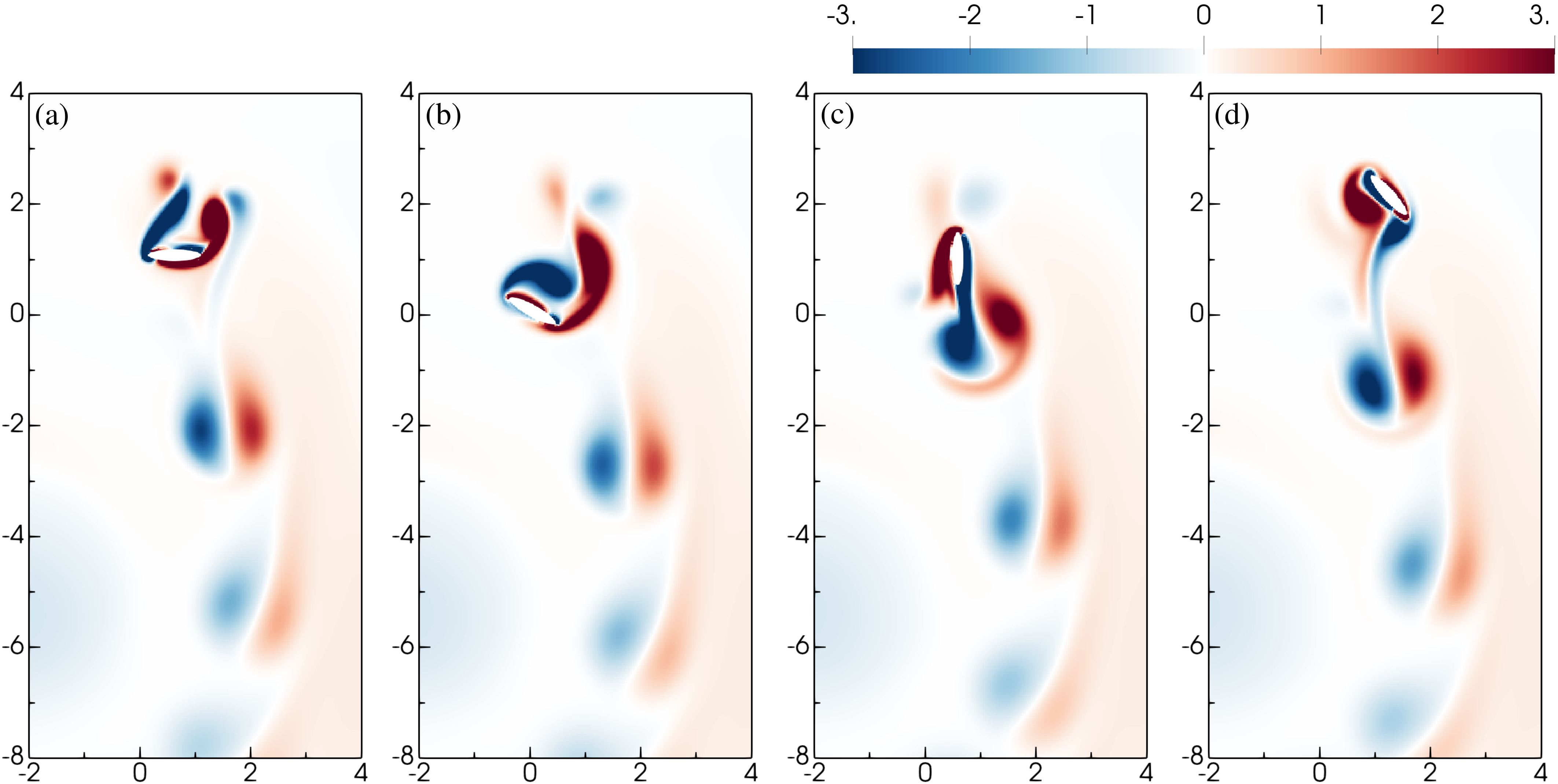}
    \caption{%
        Vorticity fields around a flapping elliptic wing at four times; (a) $t = 0.25 T$, (b) $t = 0.44 T$, (c) $t = 0.74 T$, and (d) $t = 0.99 T$.
        Because the magnitude of vorticity near the flapping wing was large, the color range was limited to $-3$ to 3 to emphasize vorticity fields far from the wing.
    }
    \label{fig_vorticity_time_variation}
  \end{center}
\end{figure}

\begin{figure}[H] 
  \begin{center}
    \includegraphics[width=.6\linewidth]{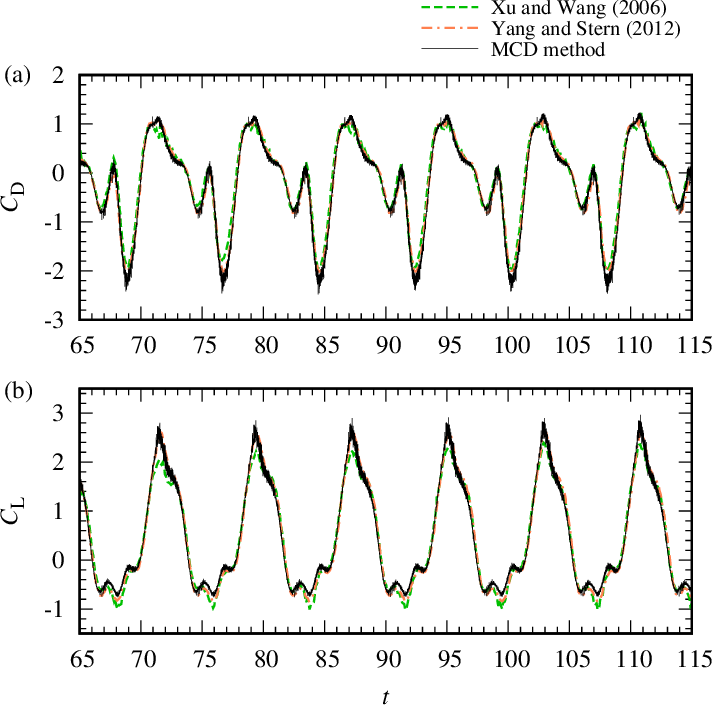}
    \caption{%
        Temporal variations of the drag coefficient $C_{\rm{D}}$ (a) and lift coefficient $C_{\rm{L}}$ (b).
        The results are shown as the green dashed lines for Xu and Wang's method \cite{xu_ImmersedInterfaceMethod_2006}, orange dash-dotted lines for Yang and Stern's method \cite{yang_SimpleEfficientDirect_2012}, and black lines for the proposed MCD method.
    }
    \label{fig_cd_cl_flapping_wing}
  \end{center}
\end{figure}

Figure \ref{fig_vorticity_time_variation} shows the vorticity fields around the flapping wing at four different times.
Compared with Xu and Wang's method \cite{xu_ImmersedInterfaceMethod_2006}, the proposed method exhibited reasonable flow fields.
The temporal changes of $C_{\rm D}$ and $C_{\rm L}$ (Figure \ref{fig_cd_cl_flapping_wing}) also achieved reasonable solutions that captured major features for the overall and peak behaviors, as shown in previous studies in which various numerical solvers are applied \cite{janewang_TwoDimensionalMechanism_2000,xu_ImmersedInterfaceMethod_2006,sui_HybridImmersedboundaryMultiblock_2007,yang_SimpleEfficientDirect_2012,cai_MovingImmersedBoundary_2017,chi_DirectionalGhostcellImmersed_2020}.

The present MCD method explicitly captures moving interfaces by DPs and solves a fluid system with a common spatial discretization, i.e., MLS reconstruction.
Thus, it does not require a particular treatment for spatial discretization near the interface, which is adopted in IBMs \cite{yang_SimpleEfficientDirect_2012, cai_MovingImmersedBoundary_2017, chi_DirectionalGhostcellImmersed_2020} and immersed interface method \cite{xu_ImmersedInterfaceMethod_2006}.
These features have an advantage in dealing with more complex and multiple boundaries.
As shown in Section \ref{two_cyl_gap}, a quadratic polynomial could be constructed even if there is only one fluid DP between the gap of two interfaces and this promises a stable and smooth solution. 
Furthermore, the common treatment of spatial discretization inside and near the interface equalizes the computational load, which is important for efficient parallel computing.


\subsection{An advanced algorithm for the DP arrangement}
As a further development, an advanced algorithm for the DP arrangement is developed.
In the original DP arrangement algorithm, the inner DPs whose masks are assigned for the inner domain are not relocated during the procedure in Eq. (\ref{eq_2nd_intermed_position_x**}), which only relocates the boundary DPs.
This procedure is updated to make the inner DPs follow the boundary surfaces as
\begin{equation}
  \mathbf{x}^{**}_{i} = 
  \left\{ \begin{array}{cl}
    \mathbf{x}^{*}_{i} + \beta \psi_{i} \hat{\mathbf{d}}_i, & \left( i \in \Lambda^{\rm I}_{i} \right), \\[2mm]
    \mathbf{x}^{*}_{i} + \psi_{i} \hat{\mathbf{d}}_i, & \left( i \in \Lambda^{\Gamma}_{i} \right).
  \end{array} \right.
  \label{eq_dp_arrng_boundary-fit_x**}
\end{equation}
Here, $\beta$ is the coefficient that controls the movement of DPs depending on the distance from the boundary surface, defined as
\begin{equation}
  \beta = 
  \left\{ \begin{array}{cl} \displaystyle
    \frac{\alpha}{2} \left[ 1 + \cos{\frac{\pi \psi_{i}}{r_{e}}} \right], & {\rm if}~ \psi_{i} < r_{e}, \\[2mm]
    0, & {\rm otherwise},
  \end{array} \right.
  \label{eq_dp_arrng_beta}
\end{equation}
where the parameter $\alpha$ determines the degree of influence of $\beta$.

\begin{figure}[H] 
  \begin{center}
    \includegraphics[width=.8\linewidth]{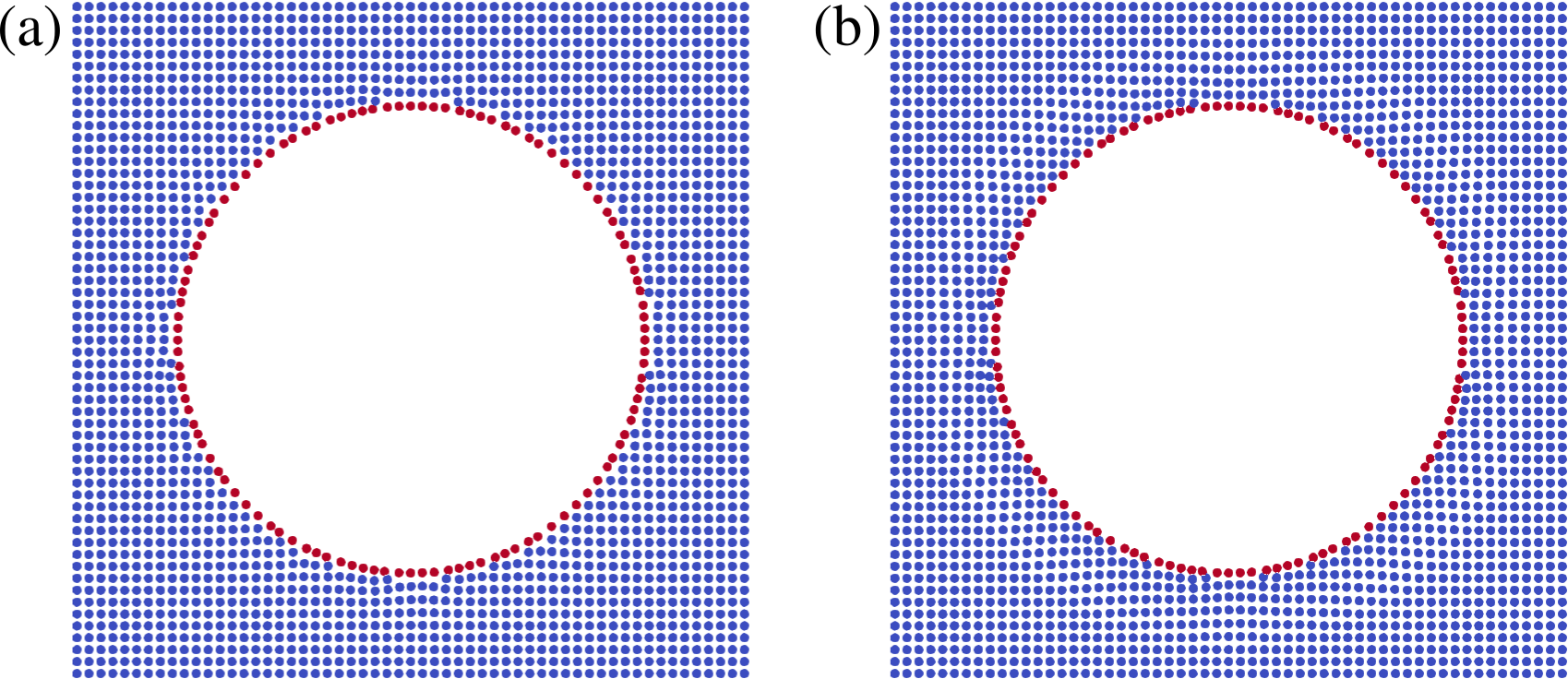}
    \caption{%
        Comparisons of the DP arrangement algorithms. (a) Original DPs ($\alpha=0$). (b) Boundary-fitted DPs ($\alpha=0.03$).
    }
    \label{fig_dp_arrng_boundary-fit}
  \end{center}
\end{figure}

Fig. \ref{fig_dp_arrng_boundary-fit} shows the DP arrangements around a circular cylinder (or disk) at $\alpha=0$ (original DPs) and $\alpha=0.03$ (boundary-fitted DPs).
Here, $r_e=5l_0$ was applied. 
Although both the DPs at $\alpha=0$ and 0.03 are distributed following the boundary surface because of the relaxation procedure, the boundary-fitted DPs at $\alpha=0.03$ are more smoothly distributed following the boundary shape.

\begin{figure}[H] 
  \begin{center}
    \includegraphics[width=0.98\linewidth]{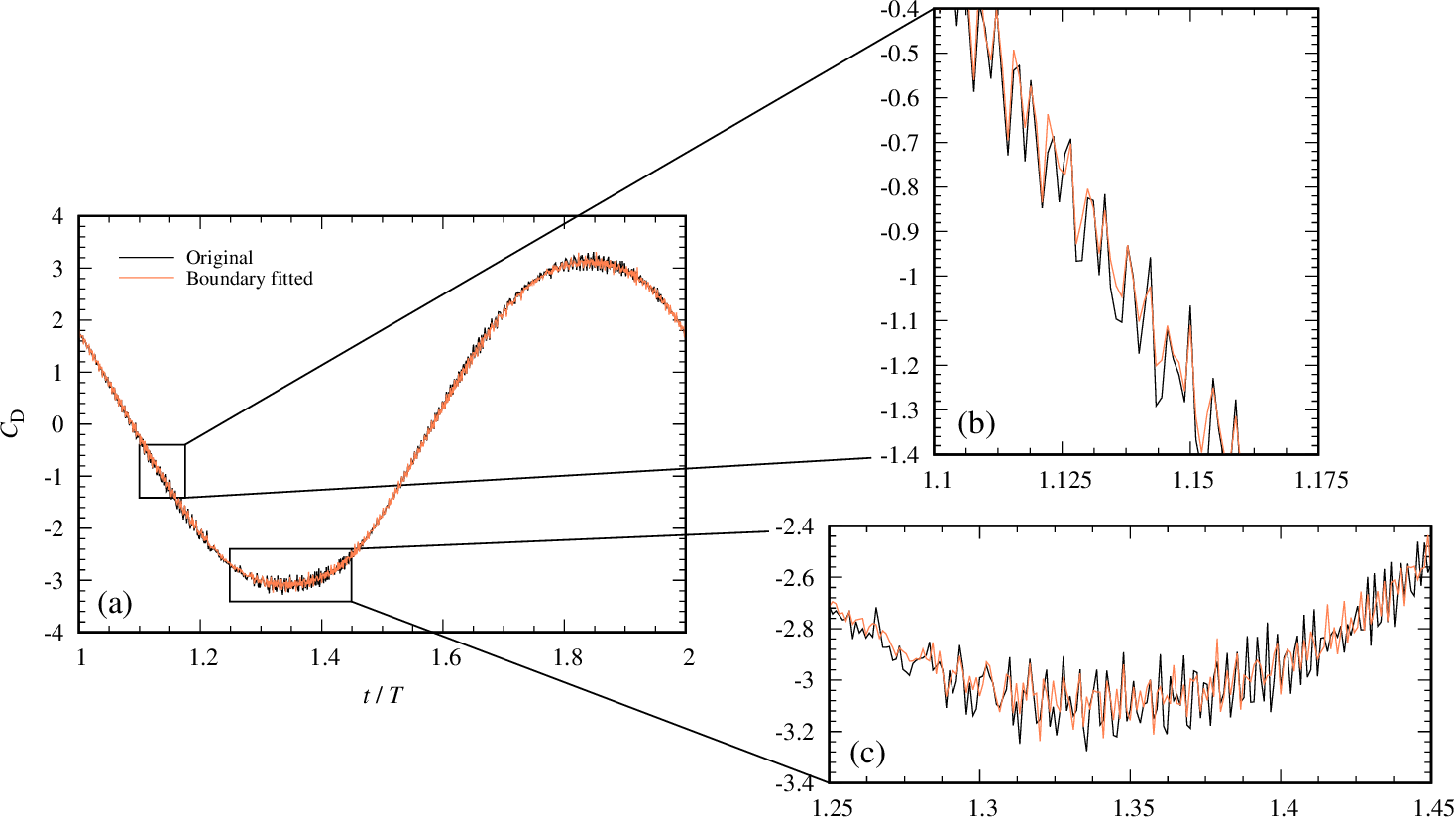}
    \caption{%
        Comparison of the time changes of the drag coefficient $C_{\rm D}$ for the single cylinder oscillation problem in one period (a), and their enlarged views (b, c).
        The orange and black lines are the results using the boundary-fitted DPs ($\alpha=0.03$) and original DPs, respectively.
    }
    \label{fig_cd_boundary-fitted_single_cyl_oscill}
  \end{center}
\end{figure}

Fig. \ref{fig_cd_boundary-fitted_single_cyl_oscill} shows the time change of the drag coefficient $C_{\rm D}$ for the single cylinder oscillation problem (shown in Section \ref{single_cyl_oscill}).
In this problem, $\alpha$ was set to 0.03.
Unphysical numerical oscillations were reduced using the boundary-fitted DPs.
It could be considered that the smooth distribution of DPs near the boundary improves smooth interpolation using the MLS approximation.

Although the current approach is not the unique choice to address the boundary fitted arrangement and needs to be further verified with some numerical examples, we believe that the MCD method has a potential to handle moving boundaries more stably due to the flexibility of the DP arrangement compared to the existing Cartesian grid-based methods including IBM.

\section{Conclusion} \label{conclusion}
In this study, a novel meshless approach for moving boundary flow simulations in the framework of the MCD method was proposed.
The proposed method represents moving boundaries by changing the role of DPs, which are constrained in the background mesh system.
Several numerical tests were conducted to validate the proposed method, and the solutions obtained agreed well with both previous numerical and experimental results.

Notable remarks are summarized as follows.
\begin{itemize}
\item The mesh constraint idea allows computational stencils to be compact and reduces the computational cost compared to wide-stencil schemes such as existing particle and meshless methods.
\item Both the fluid forces for pressure and viscous stress were evaluated accurately within a few numerical oscillations. 
\item Although numerical oscillations rose as $\Delta t$ decreased, the proposed method used the practical values for $\Delta t$ in $CFL \approx 0.2$--0.3 in moving boundary problems and obtained appropriate results that were in good agreement with previous solutions.
\item Stable and accurate calculations were possibly performed even when the gap resolution between moving boundaries was one DP.
\end{itemize}

An inherent limitation of the MCD method that the restriction of the DP movement by the background mesh would cause loss of numerical conservation and associated numerical dissipation and oscillation.
Future research should investigate how this limitation affects practical applications such as free surface flows.
Moreover, it is necessary to investigate reasons for numerical oscillations when $\Delta t$ is reduced.
Additionally, a formulation in interaction problems between a fluid and solid needs to be applied for practical problems.
It could be achieved from a numerical result that provides an accurate and less oscillating estimation of fluid forces on a moving boundary.

\section*{CRediT authorship contribution statement}
\textbf{Takeharu Matsuda:} Writing--original draft, Visualization, Validation, Software, Methodology, Investigation, Data curation, Conceptualization. \textbf{Satoshi Ii:} Writing--review \& editing, Validation, Software, Methodology, Conceptualization.

\section*{Declaration of competing interest}
The authors declare that they have no known competing financial interests or personal relationships that could have appeared to influence the work reported in this paper.

\section*{Data availability}
Data will be made available on request.

\section*{Acknowledgements}
This work was supported by JST SPRING, Grant Number JPMJSP2156; JSPS KAKENHI Grant Number JP24KJ1851, JP22K19939; the MEXT Program for Promoting Researches on the Supercomputer Fugaku (Development of human digital twins for cerebral circulation using Fugaku, JPMXP1020230118) and used computational resources of the supercomputer Fugaku provided by the RIKEN Center for Computational Science (project ID: hp230208, hp240220, hp240080).
Some computations were also carried out using the supercomputer "Flow" at the Information Technology Center, Nagoya University. 
We thank Edanz (https://jp.edanz.com/ac) for editing a draft of this manuscript.

\begin{appendix}
\renewcommand{\thesection}{Appendix \Alph{section}.}
\renewcommand{\theequation}{\Alph{section}.\arabic{equation}}  
\setcounter{equation}{0}  
\renewcommand{\thefigure}{\Alph{section}.\arabic{figure}}  
\setcounter{figure}{0}  
\section{MLS interpolation using stencils at the previous timestep} \label{appendix_A}
In this study, the information at the $k$-th step is used to approximate a physical quantity at the $k$-th step at $\mathbf{x}^{k+1}$; $\tilde{\phi}^{k} = \phi^{k} (\mathbf{x}^{k+1})$.
When the DP is in $\Omega_{\rm I}$ at $k$-th and $k+1$-th steps, $\tilde{\phi}_{c}^{k}$ is approximated by
\begin{equation}
  \tilde{\phi}_{c}^{k} = \phi_{c}^{k} + \Delta \mathbf{p}_{c}^{k} \cdot \mathbf{\Phi}_{c}^{k},
  \label{eq_appendix_A1}
\end{equation}
where $\Delta \mathbf{p}^{k} = \mathbf{p} (\Delta \mathbf{X}^{k})$, $\Delta \mathbf{X}^{k} = (\mathbf{x}^{k+1} - \mathbf{x}_{c}^{k}) / r_{s}$, and $\mathbf{\Phi}_{c}^{k}$ is the vector of polynomial coefficient at $k$-th step.

\begin{figure}[H] 
  \begin{center}
    \includegraphics[width=.85\linewidth]{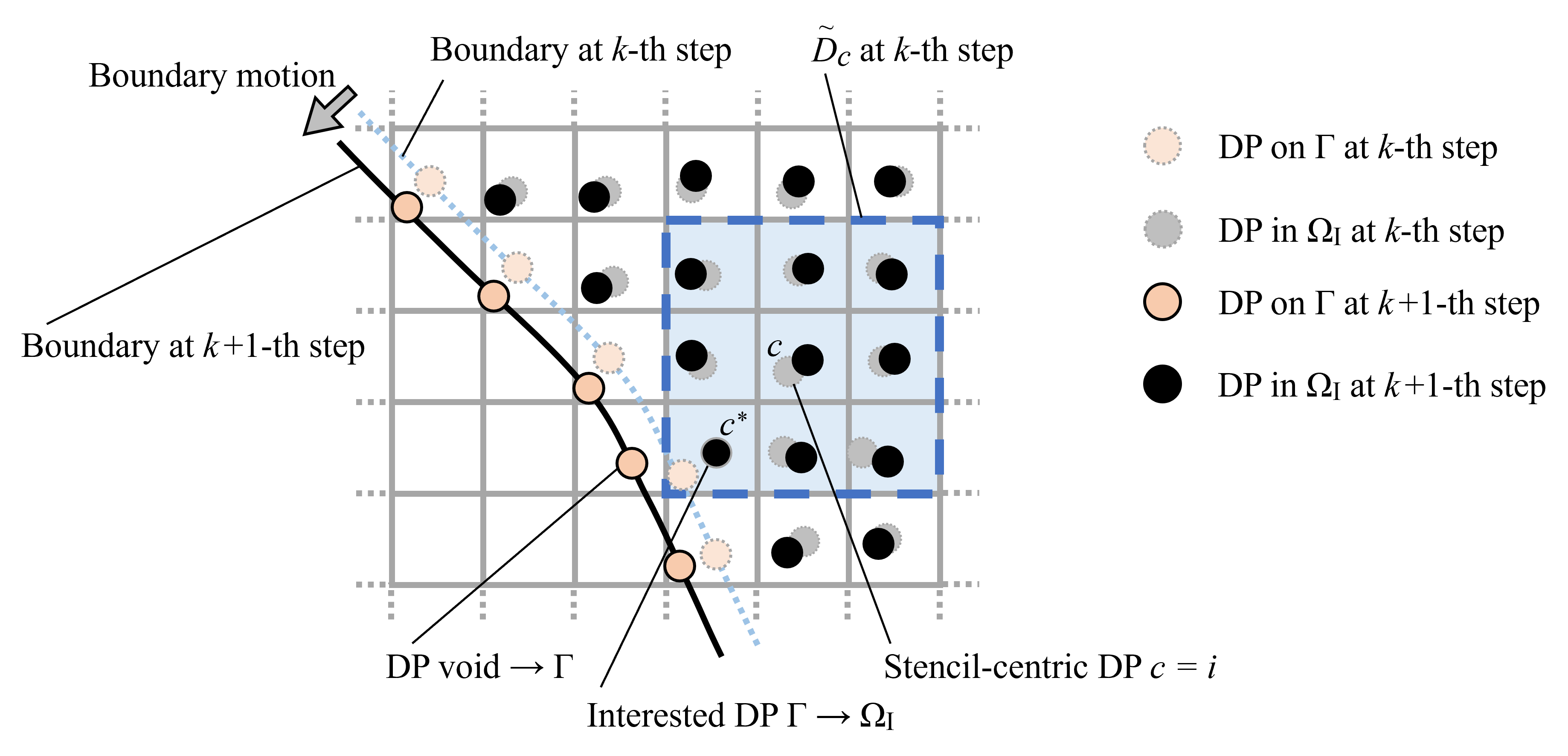}
    \caption{%
      Example of the DP rearrangement including a DP whose mask is changed to $\Gamma \rightarrow \Omega_{\rm I}$. The physical quantity $\phi_{c^{*}}^{k+1}$ on the interested $c^{*}$-th DP at $k+1$-th step is approximated by the MLS reconstruction for the stencil-centric $c$ at $k$-th step.
    }
    \label{fig_chic1_intrp_phi_changing_mask}
  \end{center}
\end{figure}

As described in Section \ref{vel_pres_cpl}, when the mask is changed to $\Gamma \rightarrow \Omega_{\rm I}$ with boundary moving, the physical information of the DP is missing.
Thus, the missing physical quantity $\tilde{\phi}^{k}$ is approximated using the parallel-shifted compact support domain, with the same idea of the interpolation on $\Gamma$ described in Section \ref{eval_fluid_forces_mov_bound} (see Figure \ref{fig_stencil_shift}).
Figure \ref{fig_chic1_intrp_phi_changing_mask} shows an example of the DP rearrangement including a DP whose mask is changed to $\Gamma \rightarrow \Omega_{\rm I}$.
The physical quantity $\tilde{\phi}_{c^{*}}^{k}$ is evaluated by
\begin{equation}
  \tilde{\phi}_{c^{*}}^{k} = \phi_{c}^{k} + \Delta \mathbf{p}_{c^{*}}^{k} \cdot \mathbf{\Phi}_{c}^{k}.
  \label{eq_appendix_A2}
\end{equation}
Note that, $\mathbf{\Phi}_{c}^{k}$ in Eq. (\ref{eq_appendix_A2}) is calculated by using the parallel-shifted compact support domain $\tilde{D}_{c}$ at $k$-th step.
\end{appendix}

\bibliographystyle{elsarticle-num}  
\bibliography{ref}  

\end{document}